\documentclass[11pt,draftcls,journal,onecolumn]{IEEEtran}
\usepackage{amsmath}
\usepackage{amssymb}
\usepackage{verbatim}
\usepackage{epsfig}
\usepackage{subfigure}
\usepackage{graphicx}
\usepackage{cases}
\usepackage{amsfonts}

\newtheorem{definition}{Definition}
\newtheorem{lemma}{Lemma}
\newtheorem{theorem}{Theorem}

\newtheorem{remark}{\indent \bf Remark}

\def\snr    {\mbox{\scriptsize\sf SNR}}
\def\Pout {P_{\textrm{out}}}

\begin{document}
\title{Performance of Hybrid-ARQ in Block-Fading Channels: A Fixed Outage Probability Analysis}

\author{Peng~Wu~and~Nihar~Jindal\thanks{The authors are with the Department of Electrical and Computer Engineering,
University of Minnesota, Minneapolis, MN 55455, USA (email: pengwu@umn.edu; nihar@umn.edu).}}
\maketitle
\vspace{-15mm}

\begin{abstract}
This paper studies the performance of \textit{hybrid}-ARQ (automatic
repeat request) in Rayleigh block-fading channels.
The long-term average transmitted rate is analyzed in a fast-fading scenario where
the transmitter only has knowledge of channel statistics, and, consistent with contemporary
wireless systems, rate adaptation is performed such that a target outage probability (after a maximum
number of H-ARQ rounds) is maintained.
H-ARQ allows for early termination once decoding is possible, and thus is a
coarse, and implicit, mechanism for rate adaptation to the instantaneous channel quality.
Although the rate with H-ARQ is not as large as the ergodic capacity,
which is achievable with rate adaptation to the instantaneous channel conditions,
even a few rounds of H-ARQ make the gap to ergodic capacity
reasonably small for operating points of interest.  Furthermore, the rate with
H-ARQ provides a significant advantage compared to systems that do not use H-ARQ and only adapt rate
based on the channel statistics.

\end{abstract}

\section{Introduction}\label{sec-intro}
ARQ (automatic repeat request) is an extremely powerful type of
feedback-based communication that is extensively
used at different layers of the network stack. The basic ARQ strategy
adheres to the pattern of transmission followed by feedback of an
ACK/NACK to indicate successful/unsuccessful decoding.
If simple ARQ or \textit{hybrid}-ARQ (H-ARQ) with Chase combining (CC) \cite{chase1972cad} is used, a NACK leads to
retransmission of the same packet in the second ARQ round.  If H-ARQ
with incremental redundancy (IR) is used,  the second transmission
is not the same as the first and instead contains some ``new''
information regarding the message (e.g., additional parity bits).
After the second round the receiver again attempts to decode, based
upon the second ARQ round alone (simple ARQ) or upon both ARQ rounds
(H-ARQ, either CC or IR). The transmitter moves on to the next
message when the receiver correctly decodes and sends back an ACK,
or a maximum number of ARQ rounds (per message) is reached.

ARQ provides an advantage by allowing for early termination once
sufficient information has been received.  As a result, it is most
useful when there is considerably uncertainty in the amount/quality
of information received.  At the network layer, this might
correspond to a setting where the network congestion is unknown to
the transmitter.  At the physical layer, which is the focus of this
paper, this corresponds to a fading channel whose
instantaneous quality is unknown to the transmitter.

Although H-ARQ is widely used in contemporary wireless systems such
as HSPA \cite{Martin}, WiMax \cite{Jeff} (IEEE 802.16e) and 3GPP LTE
\cite{Lorenzo}, the majority of research on this topic has focused on code design, e.g.,
\cite{Sesia}, \cite{Costello}, \cite{Soljanin}, while
relatively little research has focused on performance analysis of H-ARQ \cite{Lott}.
Most relevant to the present work, in \cite{CaireTuninetti} Caire and Tuninetti
established a relationship between H-ARQ throughput and mutual
information in the limit of infinite block length. For multiple
antenna systems, the diversity-multiplexing-delay tradeoff of H-ARQ
was studied by El Gamal et al. \cite{Gamal}, and the coding scheme
achieving the optimal tradeoff was introduced; Chuang et al.
\cite{Chuang} considered the optimal SNR exponent in the
block-fading MIMO (multiple-input multiple-output) H-ARQ channel with discrete input signal
constellation satisfying a short-term power constraint.  H-ARQ has
also been recently studied in quasi-static channels (i.e., the channel is
fixed over all H-ARQ rounds) \cite{Fitz}, \cite{Ravi} and shown to bring benefits to secrecy \cite{tang2007tsh}.

In this paper we build upon the results of \cite{CaireTuninetti} and
perform a mutual information-based analysis of H-ARQ in block-fading channels.
We consider a scenario where the fading is too fast to allow instantaneous channel quality
feedback to the transmitter, and thus the transmitter only has knowledge of the channel statistics,
but nonetheless each transmission experiences only a limited degree of channel selectivity.
In this setting, rate adaptation can only be performed based on channel statistics and
achieving a reasonable error/outage probability generally requires a conservative choice of rate if
H-ARQ is not used.  On the other hand, H-ARQ allows for \textit{implicit} rate adaptation to the instantaneous
channel quality because the receiver terminates transmission once the channel conditions
experienced by a codeword are good enough to allow for decoding.

We analyze the long-term average transmitted rate achieved with H-ARQ, assuming that there is
a maximum number of H-ARQ rounds and that a target outage probability at H-ARQ termination cannot be
exceeded.
We compare this rate to that achieved without H-ARQ in the same setting as well as to the ergodic
capacity, which is the achievable rate in the idealized setting where instantaneous channel information
is available to the transmitter.
The main findings of the paper are that (a) H-ARQ generally provides a significant advantage
over systems that do not use H-ARQ but have an equivalent level of channel selectivity,
(b) the H-ARQ rate is reasonably close to the ergodic capacity in many practical settings,
and (c) the rate with H-ARQ is much less sensitive to the desired outage probability than
an equivalent system that does not use H-ARQ.

%
%
%
%
%
%
%

The present work differs from prior literature in a number of important aspects.
One key distinction is that we consider systems in which the rate is adapted to the average SNR such that
a \textit{constant} target outage probability is maintained at all SNR's, whereas most prior work has considered either
fixed rate (and thus decreasing outage) \cite{Chuang} or increasing rate and \textit{decreasing} outage
as in the diversity-multiplexing tradeoff framework \cite{Gamal}\cite{Lzheng}.
The fixed outage paradigm is consistent with contemporary wireless systems where an outage
 level near 1\% is typical (see \cite{NiharAngel} for discussion), and certain conclusions depend heavily on the outage assumption.
With respect to \cite{CaireTuninetti}, note that the focus of \cite{CaireTuninetti}
is on multi-user issues,  e.g., whether or not a system becomes interference-limited
at high SNR in the regime of very large delay, whereas we consider single-user systems
and generally focus on performance with short delay constraints (i.e., maximum number of H-ARQ rounds).
In addition, we use the \textit{attempted} transmission rate, rather than the successful rate (which is used in
\cite{CaireTuninetti}), as our performance metric. This is motivated by applications such as Voice-over-IP (VoIP),
where a packet is  dropped (and never retransmitted) if it cannot be decoded after the maximum
number of H-ARQ rounds and quality of service is maintained by achieving the target (post-H-ARQ) outage
probability.  On the other hand, reliable data communication requires the use of
higher-layer retransmissions whenever H-ARQ outages occur; in such a setting, the relevant metric
is the successful rate, which is the product of the attempted transmission rate and the
success probability (i.e., one minus the post-H-ARQ outage probability).  If the post-H-ARQ outage
is fixed to some target value (e.g., $1\%$), then studying the attempted rate is effectively equivalent
to studying the successful rate.\footnote{If the post-H-ARQ outage probability can be optimized,
then a careful balancing between the attempted rate and higher-layer retransmissions should be conducted
in order to maximize the successful rate.  Although this is beyond the scope of the present paper, note that
some results in this direction can be found in \cite{wp09globecom} \cite{wesel09icc}.}

\section{System Model}\label{sec-model}

We consider a block-fading channel where the channel remains constant
over a block but varies independently from one block to another. The
$t$-th received symbol in the \textit{i}-th block is given by:
\begin{eqnarray}
y_{t,i} = \sqrt{\snr} ~ h_i x_{t,i} + z_{t,i},
\end{eqnarray}
where the index $i=1,2,\cdots$ indicates the block number, $t=1,2,\cdots, T$ indexes channel
uses within a block, $\snr$ is the average received SNR,
$h_i$ is the fading channel coefficient in the $i$-th block, and
$x_{t,i}, y_{t,i},$ and $z_{t,i}$ are the transmitted symbol, received symbol,
and additive noise, respectively.  It is assumed that $h_k$ is
complex Gaussian (circularly symmetric) with unit variance and zero mean, and that $h_1,
h_2, \ldots$ are i.i.d.. The noise $z_{t,i}$ has the same distribution as $h_k$ and is
independent across channel uses and blocks. The transmitted symbol $x_{t,i}$ is constrained to have
unit average power; we consider Gaussian inputs, and thus $x_{t,i}$ has the same distribution
as the fading and the noise.  Although we focus only on Rayleigh fading and
single antenna systems, our basic insights can be extended to
incorporate other fading distributions and MIMO as discussed
in Section \ref{sec-arq} (Remark 1).

We consider the setting where the receiver has perfect channel state
information (CSI), while the transmitter is aware of the
channel distribution but does not know the instantaneous channel
quality. This models a system in which the fading is too fast to
allow for feedback of the instantaneous channel conditions from the
receiver back to the transmitter, i.e., the channel coherence time
is not much larger than the delay in the feedback loop.
In cellular systems this is the case for moderate-to-high velocity users. This setting is often referred to as \textit{open-loop} because of the
lack of instantaneous channel tracking at the transmitter, although
other forms of feedback, such as H-ARQ, are permitted. The relevant performance metrics, notably
what we refer to as \textit{outage probability} and fixed outage transmitted rate, are
specified at the beginning of the relevant sections.

If H-ARQ is not used, we assume each codeword spans $L$ fading blocks; $L$ is therefore the
channel selectivity experienced by each codeword.  When H-ARQ is used, we make the following assumptions:
\begin{itemize}
\item The channel is constant within each H-ARQ round ($T$ symbols),
but is independent across H-ARQ rounds.\footnote{An intuitive but somewhat misleading
extension of the quasi-static fading model to the H-ARQ
setting is to assume that the channel is constant for the duration
of the H-ARQ rounds corresponding to a particular message/codeword, but is
drawn independently across different messages.  Because more H-ARQ
rounds are needed to decode when the channel quality is poor, such a
model actually changes the underlying fading distribution by
increasing the probability of poor states and reducing the
probability of good channel states. In this light, it is more accurate
model the channel across H-ARQ rounds according to a stationary and ergodic random process with a high degree
of correlation.}
\item A maximum of $M$ H-ARQ rounds are allowed.  An outage is declared if
decoding is not possible after $M$ rounds, and this outage probability
can be no larger than the constraint $\epsilon$.
\end{itemize}
Because the channel is assumed to be independent across H-ARQ rounds, $M$ is the \textit{maximum} amount of
channel selectivity experienced by a codeword. When comparing H-ARQ and no H-ARQ, we
set $L=M$ such that maximum selectivity is equalized.

It is worth noting that these assumptions on
the channel variation are quite reasonable for the fast-fading/open-loop
scenarios. Transmission slots in modern systems are typically around one millisecond, during which the
channel is roughly constant even for fast fading. \footnote{Frequency-domain channel variation within
each H-ARQ round is briefly discussed in Section \ref{sec-scalingL}.} An H-ARQ round
generally corresponds to a single transmission slot, but subsequent
ARQ rounds are separated in time by at least a few slots to
allow for decoding and ACK/NACK feedback; thus the assumption of
independent channels across H-ARQ rounds is reasonable. Moreover, a constraint on the number of H-ARQ rounds limits complexity (the decoder must retain information received in prior H-ARQ rounds in memory) and delay.


Throughout the paper we use the notation $F_{\epsilon}^{-1}(X)$ to
denote the solution $y$ to the equation $\mathbb{P}[X \leq
y]=\epsilon$, where $X$ is a random variable; this quantity is well defined wherever
it is used.


\section{Performance Without H-ARQ: Fixed-Length Coding}\label{sec-timefre}

We begin by studying the baseline scenario where H-ARQ is not used
and every codeword spans $L$ fading blocks. In this setting the
outage probability is the probability that  mutual information
received over the $L$ fading blocks is smaller than the transmitted
rate $R$ \cite[eq (5.83)]{TseVis}:
\begin{eqnarray} \label{eqoutage_time}
\Pout(R, \snr) = {\mathbb P}
\left[\frac{1}{L}\sum_{i=1}^{L}\log_2(1+\snr|h_i|^2) \leq R \right].
\end{eqnarray}
where $h_i$ is the channel in the $i$-th fading block.  The outage probability reasonably
approximates the decoding error probability for a system with strong coding
\cite{CaTaBi} \cite{FabregasCaire}, and the achievability of this error probability has
been rigorously shown in the limit of infinite block length ($T
\rightarrow \infty$) \cite{PrasadVaranasi_IT06} \cite{Malkamaki}.

Because the outage probability is a non-decreasing function of $R$,
by setting the outage probability to $\epsilon$ and solving for $R$
we get the following straightforward definition of $\epsilon$-outage capacity \cite{verdu1994gfc}:
\begin{definition}
The $\epsilon$-outage capacity with outage constraint $\epsilon$ and diversity
order $L$, denoted by $C_{\epsilon}^L(\snr)$, is the largest rate
such that the outage probability in (\ref{eqoutage_time}) is no
larger than $\epsilon$:
\begin{eqnarray}
C_{\epsilon}^L(\snr) &\triangleq& \max_{\Pout(R, \snr) \leq
\epsilon} R
\end{eqnarray}
\end{definition}
\vspace{3mm}

Using notation introduced earlier, the $\epsilon$-outage capacity can be rewritten as
\begin{eqnarray}
C_{\epsilon}^L(\snr) &=& F_{\epsilon}^{-1}\left(
\frac{1}{L}\sum_{i=1}^{L}\log_2(1+\snr|h_i|^2) \right) \\
&=& \log_2\snr +
F_{\epsilon}^{-1}\left(\frac{1}{L}\sum_{i=1}^L\log_2\left(\frac{1}{\snr}+|h_i|^2\right)\right).
\end{eqnarray}
For $L=1$, $\Pout(R, \snr)$ can be written in closed form and inverted to yield $C_{\epsilon}^{1} (\snr) =
\log_2\left(1+\log_e \left(\frac{1}{1-\epsilon}\right)\snr\right) $ \cite{TseVis}.
For $L > 1$ the outage probability cannot be written in closed form nor inverted, and therefore
$C_{\epsilon}^{L} (\snr)$ must be numerically
computed.
There are, however, two useful approximations to $\epsilon$-outage capacity.  The first one is the
high-SNR affine approximation \cite{ShamaiVerdu}, which adds a constant
rate offset term to the standard multiplexing gain characterization.
\begin{theorem}
The high-SNR affine approximation to $\epsilon$-outage capacity is given by
\begin{eqnarray}\label{theorem1}
C_{\epsilon}^L(\snr) = \log_2\snr +
F_{\epsilon}^{-1}\left(\frac{1}{L}\sum_{i=1}^L\log_2\left(|h_i|^2\right)\right) + o(1),
\end{eqnarray}
where the notation implies that the $o(1)$ term vanishes as $\snr
\rightarrow \infty$.
\end{theorem}
\begin{proof}
The proof is identical to that of the high-SNR offset characterization of MIMO channels in
\cite[Theorem 1]{prasad2005moc}, noting that single antenna block fading
is equivalent to a MIMO channel with a diagonal channel matrix.
\end{proof}

\begin{figure}
\begin{center}
\includegraphics[width = 4.5in]{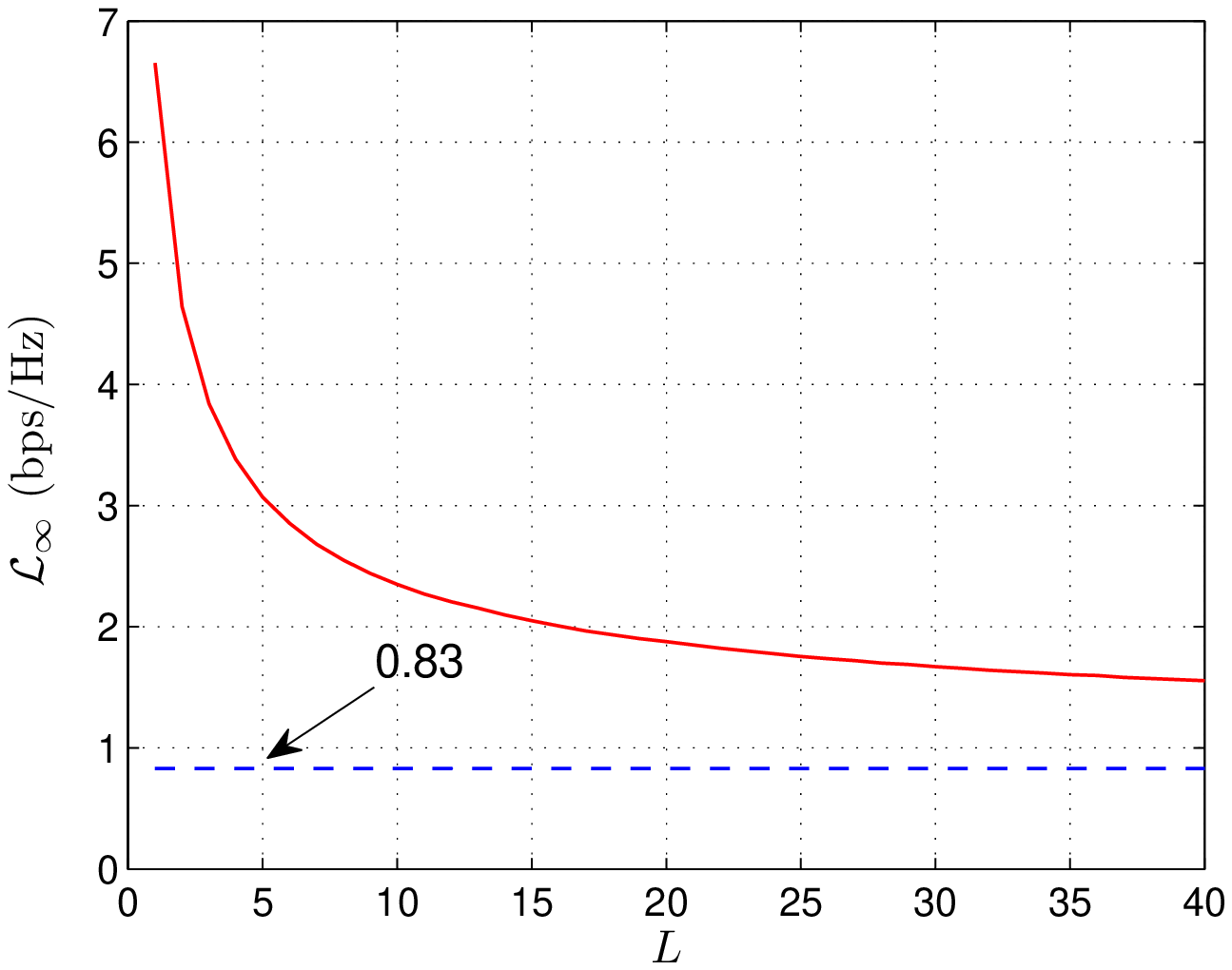}
\caption{High SNR rate offset $\mathcal{L}_{\infty}$ (bps/Hz) versus diversity order $L$ for
$\epsilon = 0.01$} \label{fig:LinfL}
\end{center}
\end{figure}

In terms of standard high SNR notation where $C(\snr) =
\mathcal{S}_{\infty}(\log_2\snr-\mathcal{L}_{\infty})+o(1)$
\cite{ShamaiVerdu}\cite{Lozano}, the multiplexing gain
$\mathcal{S}_{\infty}=1$ and the rate offset
$\mathcal{L}_{\infty} =
-F_{\epsilon}^{-1}\left(\frac{1}{L}\sum_{i=1}^L\log_2\left(|h_i|^2\right)\right)$.
The rate offset is the difference between the $\epsilon$-outage capacity and
the capacity of an AWGN channel with signal-to-noise ratio
$\snr$. Although a closed form expression for $\mathcal{L}_{\infty}$
cannot be found for $L>1$, from \cite{Salo},
\begin{eqnarray}
\mathbb{P}\left[\frac{1}{L}\sum_{i=1}^L\log_2\left(|h_i|^2\right)\leq y\right]=2^{yL}G_{1,L+1}^{L,1}\left(2^{yL}|_{0,0,\ldots,0,-1}^{0}\right),
\end{eqnarray}
where
$G_{p,q}^{m,n}\left(x|_{b_1,\ldots,b_q}^{a_1,\ldots,a_p}\right)$ is the
Meijer G-function \cite[eq. (9.301)]{Gradshteyn}.  Based on (\ref{theorem1}),
$\mathcal{L}_{\infty}$ therefore is the solution to
$2^{- \mathcal{L}_{\infty} L}G_{1,L+1}^{L,1}\left(2^{- \mathcal{L}_{\infty} L}|_{0,0,\ldots,0,-1}^{0}\right)
= \epsilon$. The rate offset $\mathcal{L}_{\infty}$ is plotted versus $L$
in Fig. \ref{fig:LinfL} for $\epsilon = 0.01$.
As $L \rightarrow
\infty$ the offset converges to $-{\mathbb E}[\log_2\left(|h|^2\right)]\approx0.83$, the offset of the ergodic Rayleigh channel \cite{Lozano}.

While the affine approximation is accurate at high SNR's, motivated by the Central Limit Theorem (CLT),
an approximation that is more accurate for moderate and low SNR's is
reached by approximating random variable
$\frac{1}{L}\sum_{i=1}^{L}\log_2(1+\snr|h_i|^2)$ by a Gaussian random variable
 with the same
mean and variance \cite{Barriac}\cite{Smith_Shafi}. The mean $\mu$ and
variance $\sigma^2$ of $\log_2(1 + \snr |h|^2)$ are given by
\cite{Alouini}\cite{McKay}:
\begin{eqnarray}\label{mean}
\mu(\snr) & = & {\mathbb E} [\log_2(1 + \snr |h|^2)] = \log_2(e) e^{1/\snr} E_1(1/\snr),\\
\sigma^2(\snr) & = &
\frac{2}{\snr}\log_2^2(e)e^{1/\snr}G_{3,4}^{4,0}\left(1/\snr|_{0,-1,-1,-1}^{0,0,0}\right)
- \mu^2(\snr), \label{variance}
\end{eqnarray}
where $E_1(x) = \int_1^{\infty} t^{-1}e^{-xt} dt$, and at high SNR the standard deviation $\sigma(\snr)$ converges to
$\frac{\pi \log_2e}{\sqrt{6}}$ \cite{McKay}.
The mutual information is thus approximated by a $\mathcal{N}(\mu(\snr),\frac{\sigma^2(\snr)}{L})$, and therefore
\begin{eqnarray}
\Pout(R, \snr) \approx Q\left(\frac{\sqrt{L}}{\sigma(\snr)}(\mu(\snr)-R)\right),
\end{eqnarray}
where $Q(\cdot)$ is the tail probability of a unit variance normal.
Setting this quantity to $\epsilon$ and then solving for $R$ yields
an $\epsilon$-outage capacity approximation \cite[eq. (26)]{Barriac}:
\begin{eqnarray} \label{Rapp}
C_{\epsilon}^L(\snr) & \approx & \mu(\snr) - \frac{\sigma(\snr)}{\sqrt{L}}Q^{-1}(\epsilon).
\end{eqnarray}

\begin{figure}
\begin{center}
\includegraphics[width = 4.5in]{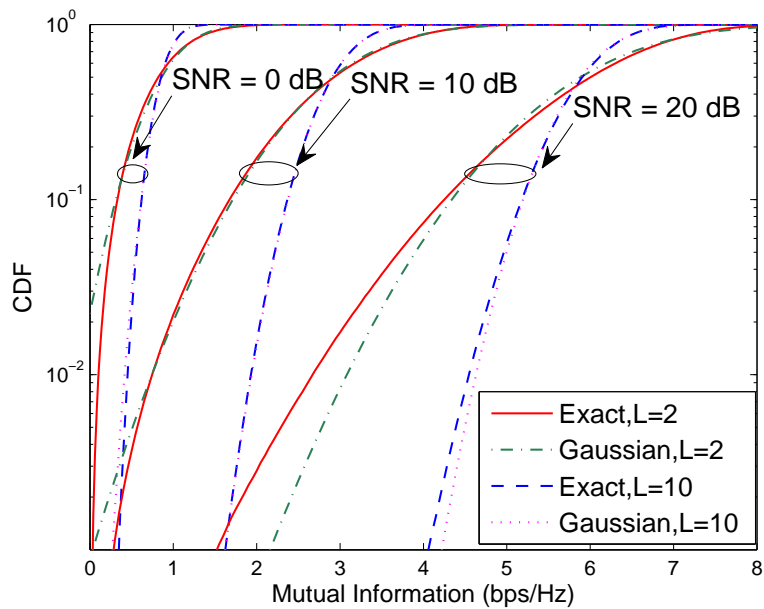}
\caption{CDF's of mutual information ($\frac{1}{2}\sum_{i=1}^2 \log_2(1 + \snr |h_i|^2)$ and $\frac{1}{10}\sum_{i=1}^{10} \log_2(1 + \snr |h_i|^2)$ respectively) for $L=2$ and $L=10$ at $\snr=0$, $10$, and $20$ dB} \label{fig:Gauss_acc}
\end{center}
\end{figure}

The accuracy of this approximation depends on how accurately the CDF
of a Gaussian matches the CDF (i.e., outage probability) of random
variable $\frac{1}{L}\sum_{i=1}^{L}\log_2(1+\snr|h_i|^2) $.  In Fig. \ref{fig:Gauss_acc} both CDF's are plotted for $L=2$ and $L=10$, and $\snr=0$, $10$, and $20$
dB.  As expected by the CLT, as $L$ increases the approximation
becomes more accurate.  Furthermore, the match is less accurate for
very small values of $\epsilon$ because the tails of the Gaussian and
the actual random variable do not precisely match.  Finally, note that the match
is not as accurate at low SNR's: this is because the mutual
information random variable has a density close to a chi-square in this regime,
and is thus not well approximated by a Gaussian.  Although not
accurate in all regimes, numerical results confirm that the Gaussian
approximation is reasonably accurate for the range of interest for
parameters (e.g., $0.01 \leq \epsilon \leq 0.2$ and $0 \leq \snr \leq 20$ dB).
More importantly, this approximation yields important insights.

\begin{figure}
\begin{center}
\includegraphics[width = 4.5in]{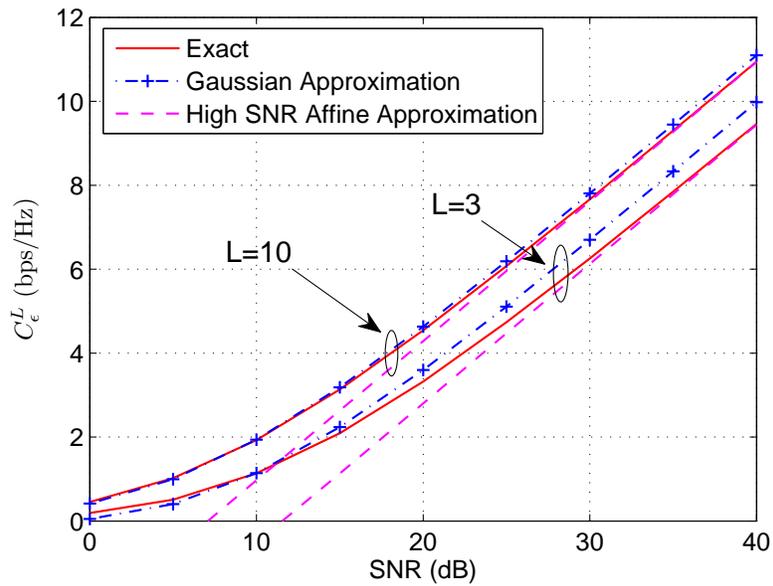}
\caption{$\epsilon$-outage capacity $C_{\epsilon}^L$ (bps/Hz) versus $\snr$ (dB) for $\epsilon =
0.01$} \label{fig:CapApp}
\end{center}
\end{figure}

In Fig. \ref{fig:CapApp} the true $\epsilon$-outage capacity $C_{\epsilon}^L(\snr)$ and
the affine and Gaussian approximations are plotted versus $\snr$ for $\epsilon = 0.01$ and
$L=3,10$. The Gaussian approximation is reasonably accurate at moderate
SNR's, and is more accurate for larger values of $L$.  On the other hand, the affine approximation,
which provides a correct high SNR offset, is asymptotically tight at high SNR.

\subsection{Ergodic Capacity Gap} \label{section-gap}
When evaluating the effect of the diversity order $L$, it is useful to compare
the ergodic capacity $\mu(\snr)$ and $C_{\epsilon}^L(\snr)$. By Chebyshev's inequality, for any $0<w<\mu$,
\begin{eqnarray}
\mathbb{P}\left[\left| \mu(\snr)- \frac{1}{L}\sum_{i=1}^{L}\log_2(1+\snr|h_i|^2) \right| \geq w\right]\leq \frac{\sigma^2 (\snr)}{L w^2}
\end{eqnarray}
By replacing $w$ with $\mu(\snr) - R$ and equating the right hand side (RHS) with $\epsilon$, we get
\begin{eqnarray} \label{C_eps_lower}
\mu(\snr) - \frac{\sigma(\snr)}{\sqrt{L\epsilon}} \leq C_{\epsilon}^L(\snr) \leq
\mu(\snr) + \frac{\sigma(\snr)}{\sqrt{L\epsilon}}.
\end{eqnarray}
This implies $C_{\epsilon}^L(\snr) \rightarrow \mu(\snr)$ as $L \rightarrow \infty$, as intuitively expected;
reasonable values of $\epsilon$ are smaller than $0.5$, and thus we expect convergence to occur from below.

\begin{figure}
\begin{center}
\includegraphics[width = 4.5in]{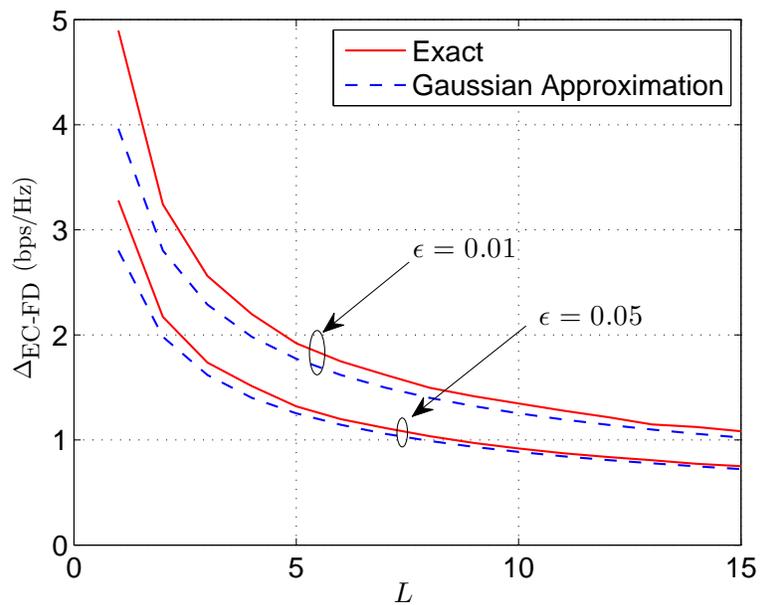}
\caption{Ergodic capacity - $\epsilon$-outage capacity difference $\Delta_{\textrm{EC - FD}}$ (bps/Hz) versus diversity order L,\ at $\snr = 20$ dB}
\label{fig:gapfreq}
\end{center}
\end{figure}

In order to capture the speed at which this convergence occurs, we define the quantity
$\Delta_{\rm EC - FD}$ as the difference between the ergodic and $\epsilon$-outage capacities.
Based on (\ref{C_eps_lower}) we can upper bound $\Delta_{\rm EC - FD}$ as:
\begin{eqnarray}
\Delta_{\rm EC - FD}(\snr) = \mu(\snr) - C_{\epsilon}^L(\snr)\leq \frac{\sigma(\snr)}{\sqrt{L\epsilon}}.
\end{eqnarray}
This bound shows that the rate gap goes to zero at least as fast as $O \left(1/\sqrt{L}\right)$.
Although we cannot rigorously claim that $\Delta_{\rm EC - FD}$ is of order $1/\sqrt{L}$, by (\ref{Rapp})
the Gaussian approximation to this quantity is:
\begin{eqnarray}\label{offset}
\Delta_{\rm EC - FD}(\snr) \approx
\frac{\sigma(\snr)}{\sqrt{L}}Q^{-1}(\epsilon),
\end{eqnarray}
which is also $O \left(1/\sqrt{L}\right)$. This approximation becomes more accurate
as $L \rightarrow \infty$, by the CLT,  and thus is expected to correctly capture the scaling with $L$.
Note that (\ref{offset}) has the interpretation that the rate must be
$\frac{Q^{-1}(\epsilon)}{\sqrt{L}}$ deviations below the ergodic capacity $\mu(\snr)$
in order to ensure $1-\epsilon$ reliability. In Fig. \ref{fig:gapfreq}
the actual capacity gap and the approximation in (\ref{offset}) are plotted for
$\epsilon = 0.01$ and $\epsilon = 0.05$ with $\snr = 20$ dB,
and a reasonable match between the approximation and the exact gap is seen.

\section{Performance with Hybrid-ARQ} \label{sec-arq}
We now move on to the analysis of hybrid-ARQ, which will be shown to
provide a significant performance advantage relative to the baseline of
non-H-ARQ performance. H-ARQ is clearly a variable-length code, in which case the average transmission
rate must be suitably defined. If each message contains $b$ information bits and each
ARQ round corresponds to $T$ channel symbols, then the \textit{initial transmission rate} is
$R_{\textrm{init}} \triangleq \frac{b}{T}$ bits/symbol. If random variable $X_i$ denotes
the number of H-ARQ rounds used for the $i$-th message, then a total of
$\sum_{i=1}^N X_i$ H-ARQ rounds are used and the average transmission rate (in bits/symbol or bps/Hz)
across those $N$ messages is:
\begin{equation}
\frac{N b} {T \sum_{i=1}^N X_i} = \frac{ R_{\textrm{init}} }{ \frac{1}{N} \sum_{i=1}^N X_i}.
\end{equation}
We are interested in the long-term average transmission rate, i.e., the case where $N \rightarrow \infty$.
By the law of large numbers (note that the $X_i$'s are i.i.d. in our model),
$\frac{1}{N} \sum_{i=1}^N X_i \rightarrow {\mathbb E}[X]$ and thus the rate converges to
\begin{equation} \label{eq-defn_arq_rate}
\frac{R_{\textrm{init}}}{{\mathbb E}[X]} ~~ \textrm{bits/symbol}
\end{equation}
Here $X$ is the random variable representing the number of H-ARQ rounds per message; this random
variable is determined by the specifics of the H-ARQ protocol.




In the remainder of the paper we focus on incremental redundancy (IR) H-ARQ because it is the most
powerful type of H-ARQ, although we compare IR to Chase combining in Section \ref{sec-chase}.
In \cite{CaireTuninetti} it is
shown that mutual information is \textit{accumulated} over H-ARQ
rounds when IR is used, and that decoding is possible once the
accumulated mutual information is larger than the number of
information bits in the message. Therefore, the number of H-ARQ
rounds $X$ is the smallest number $m$ such that:
\begin{equation}
\sum_{i=1}^m \log_2(1+ \snr |h_i|^2) > R_{\textrm{init}}.
\end{equation}
The number of rounds is upper bounded by $M$, and an outage occurs
whenever the mutual information after $M$ rounds is smaller than $R_{\textrm{init}}$:
\begin{eqnarray} \label{eq-ir_outgae}
\Pout^{\textrm{IR},M}(R_{\textrm{init}}) = { \mathbb P} \left[\sum_{i=1}^M \log_2(1+ \snr |h_i|^2) \leq R_{\textrm{init}} \right].
\end{eqnarray}
This is the same as the expression for outage
probability of $M$-order diversity without H-ARQ in
(\ref{eqoutage_time}), except that mutual
information is summed rather than averaged over the $M$ rounds. This
difference is a consequence of the fact that $R_{\textrm{init}}$
is defined for transmission over one round rather than all $M$
rounds; dividing by ${\mathbb E}[X]$ in (\ref{eq-defn_arq_rate}) to obtain the
average transmitted rate makes the expressions consistent.
Due to this relationship, if the initial rate is set as
$R_{\textrm{init}} = M \cdot C_{\epsilon}^M$, where $ C_{\epsilon}^M$ is the $\epsilon$-outage capacity for $M$-order diversity without H-ARQ, then the outage at H-ARQ termination
is $\epsilon$.

In order to simplify expressions, it is useful to define
$A_k(R_{\textrm{init}})$ as the probability that the accumulated mutual information
after $k$ rounds is smaller than $R_{\textrm{init}}$:
\begin{equation}
A_k(R_{\textrm{init}}) \triangleq { \mathbb P} \left[\sum_{i=1}^k \log_2(1+\snr |h_i|^2) \leq R_{\textrm{init}} \right].
\end{equation}
The expected number of H-ARQ rounds per message is therefore given by:
\begin{eqnarray} \label{ex}
{\mathbb E}[X] = 1 + \sum_{k=1}^{M-1}\mathbb{P}[X>k] = 1+\sum_{k=1}^{M-1}A_k(R_{\textrm{init}}).
\end{eqnarray}

The long-term average transmitted rate, which is denoted as $C_{\epsilon}^{\textrm{IR},M}$,
is defined by (\ref{eq-defn_arq_rate}).  With initial rate $R_{\textrm{init}} = M \cdot C_{\epsilon}^M$
we have:\footnote{All quantities in this expression except $M$ are
actually functions of $\snr$.  For the sake of compactness, however,
dependence upon $\snr$ is suppressed in this and subsequent
expressions, except where explicit notation is necessary.}
\begin{eqnarray} \label{IR_cap}
C_{\epsilon}^{\textrm{IR},M} \triangleq \frac{R_{\textrm{init}}}{\mathbb{E}[X]} = \left( \frac{M}{{\mathbb E}[X]} \right) C_{\epsilon}^M.
\end{eqnarray}
Note that $C_{\epsilon}^{\textrm{IR},M}$ is the \textit{attempted} long-term average transmission rate, as
discussed in Section \ref{sec-intro}.  For the sake of brevity this quantity is referred to as the H-ARQ rate; this is
not to be confused with the initial rate $R_{\textrm{init}}$. Similarly, we refer to $\epsilon$-outage capacity $C_{\epsilon}^M$ as the non-H-ARQ rate in the rest of the paper.

Because ${\mathbb E}[X] \leq M$, the H-ARQ rate is at least as large as the
non-H-ARQ rate, i.e., $C_{\epsilon}^{\textrm{IR},M} \geq C_{\epsilon}^{M}$,
and the advantage with respect to the non-H-ARQ benchmark is precisely
the multiplicative factor $\frac{M}{{\mathbb E}[X]}$. This difference is explained
as follows.  Because $R_{\textrm{init}} = M \cdot C_{\epsilon}^M$, each message/packet contains $ C_{\epsilon}^M M
T$ information bits regardless of whether H-ARQ is used. Without H-ARQ
these bits are always transmitted over $MT$ symbols, whereas with
H-ARQ an average of only ${\mathbb E}[X] T$ symbols are required.

\begin{figure}
\begin{center}
\includegraphics[width = 4.5in]{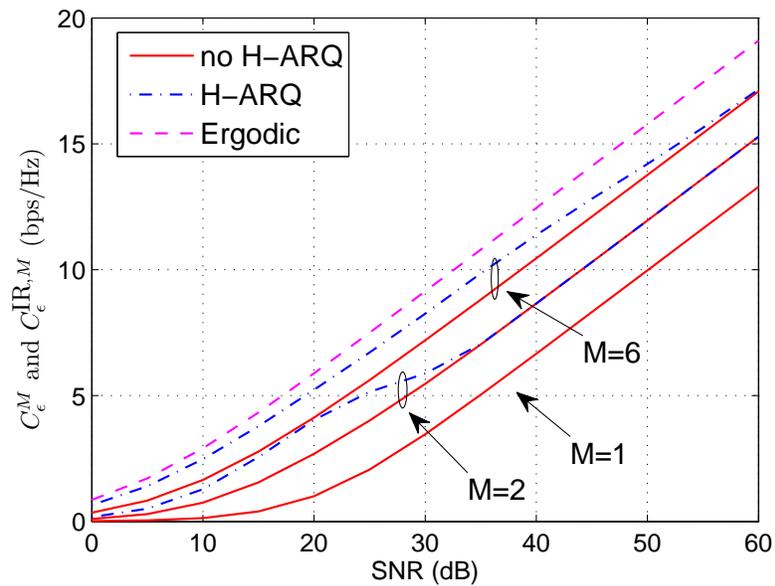}
\caption{Ergodic capacity, H-ARQ rate, and non-H-ARQ rate (bps/Hz) versus $\snr$ (dB) for $\epsilon = 0.01$} \label{fig:IRARQnARQ}
\end{center}
\end{figure}

In Fig. \ref{fig:IRARQnARQ} the average
 rates with ($C_{\epsilon}^{\textrm{IR},M}$) and without H-ARQ
($C_{\epsilon}^{M}$) are plotted versus $\snr$ for $\epsilon = 0.01$ and $M=1,2$
and $6$ ($M=1$ does not allow for H-ARQ in our model). Ergodic capacity is also
plotted as a reference. Based on the figure, we immediately
notice:
\begin{itemize}
\item H-ARQ with 6 rounds outperforms H-ARQ with 2 rounds.
\item H-ARQ provides a significant advantage relative to non-H-ARQ for the same value of $M$ for a wide range of SNR's, but this advantage
vanishes at high SNR.
\end{itemize}
Increasing rate with $M$ is to be expected, because larger $M$ corresponds to more time diversity
and more early termination opportunities.  The behavior with respect
to SNR is perhaps less intuitive.
The remainder of this section is devoted to quantifying and
explaining the behavior seen in Fig. \ref{fig:IRARQnARQ}. We begin
by extending the Gaussian approximation to H-ARQ, then examine
performance scaling with respect to $M$, $\snr$, and $\epsilon$, and finally compare
IR to Chase Combining.


\subsection{Gaussian Approximation} \label{sec-arq-gaussian}

By the definition of $A_k(\cdot)$ and (\ref{ex})-(\ref{IR_cap}),
the H-ARQ rate can be written as:
\begin{eqnarray}  \label{IR_cap2}
C_{\epsilon}^{\textrm{IR},M} & = &
\frac{A_M^{-1}(\epsilon)}{1+\sum_{k=1}^{M-1}A_k\left(A_M^{-1}(\epsilon)\right)},
\end{eqnarray}
where $A_M^{-1}(\cdot)$ refers to the inverse of function
$A_M(\cdot)$.  If we use the approach of Section \ref{sec-timefre}
and approximate the mutual information accumulated in $k$ rounds by
a Gaussian with mean $\mu k$ and variance $\sigma^2 k$, where $\mu$
and $\sigma^2$ are defined in (\ref{mean}) and (\ref{variance}), we
have:
\begin{eqnarray}\label{A_approx}
A_k(R_{\textrm{init}}) \approx Q \left( \frac{  \mu k - R_{\textrm{init}}}{ \sigma \sqrt{k} } \right).
\end{eqnarray}
Similar to (\ref{Rapp}), the initial rate
$R_{\textrm{init}}=A_M^{-1}(\epsilon)$ can be approximated as $M \left[\mu -
\frac{\sigma}{\sqrt{M}} Q^{-1} (\epsilon) \right]$.  Applying the approximation
of $A_k(R_{\textrm{init}})$ to each term in (\ref{IR_cap2}) and using the property $1-Q(x) = Q(-x)$
yields:
\begin{eqnarray}\label{IR_cap_Gaussian}
C_{\epsilon}^{\textrm{IR},M}
\approx  \frac{M \left[\mu -
\frac{\sigma}{\sqrt{M}}Q^{-1}(\epsilon)\right]}{M - \sum_{k=1}^{M-1}
Q\left(\frac{M-k}{\sqrt{k}}\frac{\mu}{\sigma}-\sqrt{\frac{M}{k}}Q^{-1}(\epsilon)\right)}.
\end{eqnarray}
This approximation is easier to compute than the actual H-ARQ rate and
is reasonably accurate.
Furthermore, it is useful for the insights it can provide.

\subsection{Scaling with H-ARQ Rounds $M$} \label{sec-scalingL}
In this section we study the dependence of the H-ARQ rate on $M$.
We first show convergence to the ergodic capacity as $M\rightarrow \infty$:
\begin{theorem} \label{thm-H-ARQ_converge}
For any $\snr$,  the H-ARQ rate converges to the ergodic capacity as $M\rightarrow \infty$:
\begin{eqnarray}
& &\lim_{M\rightarrow \infty} C_{\epsilon}^{\textrm{IR},M}(\snr)  =  \mu(\snr) \label{thm_scaling_M}
\end{eqnarray}
\end{theorem}
\begin{proof}
See Appendix A.
\end{proof}
To quantify how fast this convergence is, similar to Section \ref{section-gap} we investigate the difference between the ergodic
capacity and the H-ARQ rate.  Defining
$\Delta_{\rm EC-IR} \triangleq \mu(\snr) - C_{\epsilon}^{\textrm{IR},M}(\snr)$ we have
\begin{eqnarray}
\Delta_{\rm EC-IR}
&=& \frac{\mu}{{\mathbb E}[X]} \left( {\mathbb E}[X] - \frac{M \cdot C_{\epsilon}^M}{\mu} \right)\approx \frac{\mu}{{\mathbb E}[X]} \left( {\mathbb E}[X] - \left( M - \frac{\sigma}{\mu} Q^{-1}(\epsilon) \sqrt{M} \right) \right)\label{gap_dif}
\end{eqnarray}
where the approximation follows from $C_{\epsilon}^M \approx \mu - \frac{\sigma Q^{-1}(\epsilon)}{ \sqrt{M}}$ in (\ref{Rapp}).
Because ${\mathbb E}[X]$ is on the order of $M$ (as established in the proof of Theorem \ref{thm-H-ARQ_converge}),
the key is the behavior of the term
${\mathbb E}[X] - \left( M - \frac{\sigma}{\mu} Q^{-1}(\epsilon) \sqrt{M} \right)$.

To better understand ${\mathbb E}[X]$ we again return to the Gaussian approximation.
While the CDF of $X$ is defined by $\mathbb{P} \left( X \leq k \right) =1- A_k(R_{\textrm{init}})$ (for $k =1, \ldots, M-1$), we use
$\tilde{X}$ to denote the random variable using the Gaussian approximation and thus define its CDF (for integers $k$) as:
\begin{eqnarray}
\mathbb{P} \left( \tilde{X} \leq k \right) = Q \left( \frac{\mu (M - k)-\sqrt{M} \sigma Q^{-1}(\epsilon) }{\sigma \sqrt{k}} \right)
\end{eqnarray}
where we have used $A_k(R_{\textrm{init}}) \approx Q \left( \frac{  \mu k - R_{\textrm{init}}}{ \sigma \sqrt{k} } \right)$
evaluated with $R_{\textrm{init}} = M C_{\epsilon}^M \approx M \mu - \sqrt{M} \sigma Q^{-1}(\epsilon)$.
From this expression, we can immediately see that
the {\em median} of $\tilde{X}$ is
$\left\lceil M-\frac{\sigma}{\mu}Q^{-1}(\epsilon)\sqrt{M}\right\rceil$ \cite{Fristedt}.  If this was equal to the
mean of $X$, then by (\ref{gap_dif})
the rate difference would be well approximated by $\frac{\beta \mu}{M}$, where $\beta$ is the
difference between $\left\lceil M-\frac{\sigma}{\mu}Q^{-1}(\epsilon)\sqrt{M}\right\rceil$ and
$\left( M-\frac{\beta}{\mu}Q^{-1}(\epsilon)\sqrt{M}\right)$ and thus is no larger than one.
By studying the characteristics of $\tilde{X}$ (and of $X$) we can see that the median is in fact quite close to the mean.
A tedious calculation  in Appendix B gives the following approximation to $\mathbb{E}[X]$:
\begin{eqnarray}\label{app_Ex}
{\mathbb E}[X]& \approx &
M-\frac{\sigma}{\mu}Q^{-1}(\epsilon)\sqrt{M} +0.5(1-\epsilon) -\frac{\sigma}{\mu}\sqrt{M}\int_{Q^{-1}(\epsilon)}^{\infty}Q(x)dx,
\end{eqnarray}
which is reasonably accurate for large $M$.
The most important factor is the term $0.5(1-\epsilon)$, which is due to the fact that only
an integer number of H-ARQ rounds can be used.
The factor $-\frac{\sigma}{\mu}\sqrt{M}\int_{Q^{-1}(\epsilon)}^{\infty}Q(x)dx$ exists because the random variable
is truncated at the point where its CDF is $1 - \epsilon$.

\begin{figure}
\begin{center}
\includegraphics[width = 4.5in]{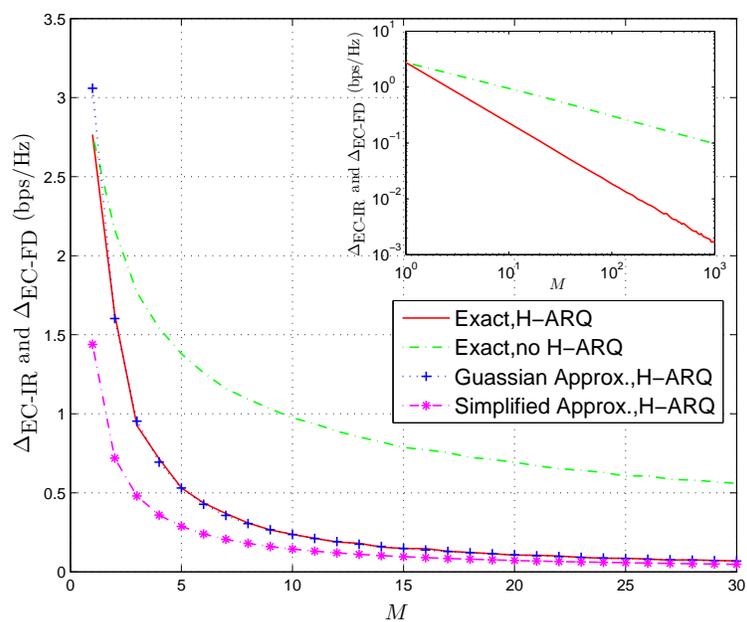}
\caption{Ergodic capacity - H-ARQ rate difference $\Delta_{\textrm{EC-IR}}$ (bps/Hz) and ergodic capacity - non-H-ARQ rate difference $\Delta_{\textrm{EC-FD}}$ (bps/Hz) versus $M$ for $\epsilon=0.01$ at $\snr=10$ dB} \label{GapIR}
\end{center}
\end{figure}

Applying this into (\ref{gap_dif}), the rate difference can be approximated as:
\begin{eqnarray}\label{app_gap}
\Delta_{\rm EC-IR}& \approx &
\frac{\mu\left(0.5(1-\epsilon)-\frac{\sigma}{\mu}\sqrt{M}\int_{Q^{-1}(\epsilon)}^{\infty}Q(x)dx\right)}{M-\frac{\sigma}{\mu}Q^{-1}(\epsilon)\sqrt{M}-\frac{\sigma}{\mu}\sqrt{M}\int_{Q^{-1}(\epsilon)}^{\infty}Q(x)dx+0.5(1-\epsilon)}.
\end{eqnarray}
The denominator increases with $M$ at the order of $M$ (more precisely as $M - \sqrt{M}$), while the
numerator actually decreases with $M$ and can even become negative if $M$ is extremely large.
For reasonable values of $M$, however, the negative term in the numerator is essentially
inconsequential (for example, if $\epsilon=0.01$ and $\snr=10$ dB, the negative term is much smaller
than $0.5(1-\epsilon)$ for $M<5000$) and thus can be reasonably neglected.  By ignoring
this negative term and replacing the denominator with the leading order $M$ term, we
get a further approximation of the rate gap:
\begin{eqnarray}\label{app_gap2}
\Delta_{\rm EC-IR}& \approx & \frac{0.5 (1 - \epsilon) \mu}{M}
\end{eqnarray}
Based on this approximation, we see that the rate gap decreases roughly on
the order $O\left(1/M\right)$, rather than the $O\left(1/\sqrt{M}\right)$ decrease without H-ARQ. In Fig. \ref{GapIR} we plot the exact capacity gap with and without H-ARQ, as well as the Gaussian approximation to the H-ARQ gap (\ref{app_gap}) and its simplified form in (\ref{app_gap2}) for $\epsilon = 0.01$ at $\snr = 10$ dB.
Both approximations are seen to be reasonably accurate especially for large $M$.
In the inset plot, which is in log-log scale, we see that the exact capacity gap goes to zero
at order $1/M$, consistent with the result obtained from our approximation.

The fast convergence with H-ARQ can be intuitively explained as follows. If transmission could be
stopped precisely when enough mutual information has been received, the transmitted
rate would be exactly matched to the instantaneous mutual information and thus ergodic capacity
would be achieved.  When H-ARQ is used, however, transmission can only be
terminated at the end of a round as, opposed to within a round, and thus a small amount of the
transmission can be wasted.  This "rounding error", which is reflected in the
$0.5(1-\epsilon)$ term in (\ref{app_gap}) and (\ref{app_gap2}), is essentially the only
penalty incurred by using H-ARQ rather than explicit rate adaptation.


\begin{remark}
Because the value of H-ARQ  depends primarily on the mean and
variance of the mutual information in each H-ARQ round, our basic insights can
be extended to multiple-antenna channels and to channels with frequency (or time) diversity within each
ARQ round if the change in mean and variance is accounted for.  For example, with order $F$ frequency diversity
the mutual information in the $i$-th H-ARQ round becomes $\frac{1}{F}\sum_{l=1}^F \log(1 + \snr |h_{i,l}|^2)$,
where $h_{i,l}$ is the channel in the $i$-th round on the $l$-th frequency channel.  The mean mutual
information is unaffected, while the variance is decreased by a factor of $\frac{1}{F}$.
\hfill $\lozenge$
\end{remark}

\subsection{Scaling with SNR} \label{sec-snr}

In this section we quantify the behavior of H-ARQ as a function of the average SNR.\footnote{
Because constant outage corresponds to the full-multiplexing point, the results of
\cite{Gamal} imply that $C_{\epsilon}^{\textrm{IR},M}$ cannot have a multiplexing gain/pre-log larger than one (in \cite{Gamal} it is shown that H-ARQ does not
increase the full multiplexing point).  However, the DMT-based results of \cite{Gamal} do not provide
rate-offset characterization as in Theorems \ref{thm-high} and \ref{thm-high2}.}
Fig. \ref{fig:IRARQnARQ} indicated that
the benefit of H-ARQ vanishes at high SNR, and the following theorem makes this precise:
\begin{theorem} \label{thm-high}
If $\snr$ is taken to infinity while keeping $M$ fixed, the expected number of H-ARQ rounds converges to $M$ and
 the H-ARQ rate converges to $C_{\epsilon}^M(\snr)$, the non-H-ARQ rate with the same selectivity:
\begin{eqnarray}
& &\lim_{\snr\rightarrow \infty} {\mathbb E}[X]  =  M\\ \label{eq-snr1}&
&\lim_{\snr\rightarrow \infty} \left[C_{\epsilon}^{\textrm{IR},M}(\snr) -
C_{\epsilon}^M(\snr)\right]  =  0 \label{eq-snr2}
\end{eqnarray}
\end{theorem}

\begin{proof}
See Appendix C.
\end{proof}

\begin{figure}
\begin{center}
\includegraphics[width = 4.5in]{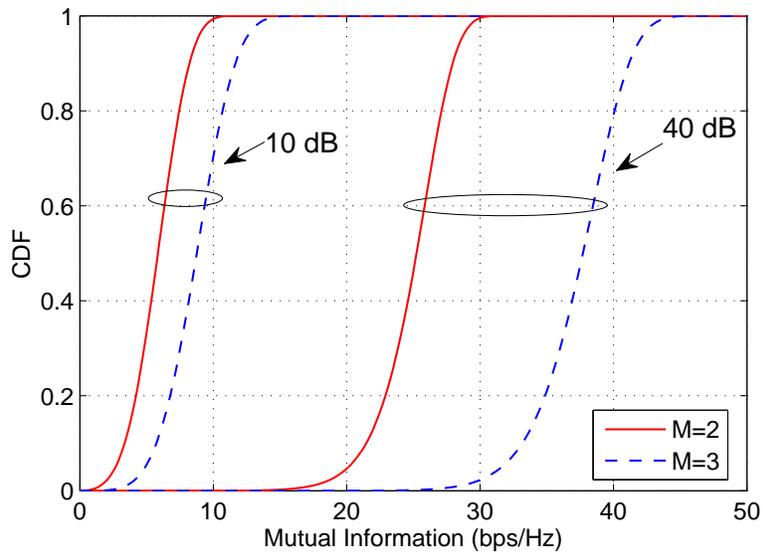}
\caption{CDF's of accumulated mutual information over $2$ and $3$ rounds
(random variables $\sum_{i=1}^2 \log_2(1 + \snr |h_i|^2)$ and $\sum_{i=1}^3 \log_2(1 + \snr |h_i|^2)$, respectively)
at $\snr = 10$ and  $40$ dB} \label{fig:CDFvsMul}
\end{center}
\end{figure}

The intuition behind this result can be gathered from Fig. \ref{fig:CDFvsMul}, where the CDF's of the accumulated mutual information after $2$ and $3$ rounds are plotted for
for $\snr = 10$ and $40$ dB.  If $M=3$ the initial rate is set at the $\epsilon$-point of the CDF of
$\sum_{i=1}^3 \log(1 + \snr |h_i|^2)$. Because the CDF's overlap for $M=2$ and $M=3$ considerably when $\snr=10$ dB, there is a
large probability that sufficient mutual information is accumulated after $2$ rounds and thus early termination occurs.
However, the overlap between these CDF's disappears as $\snr$ increases, because
$\sum_{i=1}^k \log(1 + \snr |h_i|^2) \approx k \log \snr + \sum_{i=1}^k \log(|h_i|^2)$,
and thus the early termination probability vanishes.


Although the H-ARQ advantage eventually vanishes, the advantage persists
throughout a large SNR range and the Gaussian approximation (Section \ref{sec-arq-gaussian}) can be
used to quantify this.  The probability of terminating in strictly less than $M$ rounds is approximated by:
\begin{eqnarray}
\mathbb{P}[X \leq M-1] & \approx &
Q \left( \frac{ \mu(\snr) - \sqrt{M} \sigma(\snr) Q^{-1}(\epsilon)}{\sigma \sqrt{M-1}} \right)
\end{eqnarray}
In order for this approximation to be greater than one-half we require the numerator inside the $Q$-function to be less than zero, which
corresponds to
\begin{eqnarray}
\frac{\mu(\snr)}{\sigma(\snr)} \leq \sqrt{M} Q^{-1}(\epsilon)  ~~~ \textrm{or} ~~~
\sqrt{M} \geq \frac{\mu(\snr)}{\sigma(\snr) Q^{-1}(\epsilon)} \label{eq-gauss_approx_high}
\end{eqnarray}
As $\snr$ increases  $\mu(\snr)$ increases without bound whereas $\sigma(\snr)$ converges to a
constant.  Thus $\mu(\snr) / \sigma(\snr) $ increases quickly with $\snr$, which makes the probability of early termination
vanish.  From this we see that the H-ARQ advantage lasts longer (in terms of $\snr$) when $M$ is larger.
The second inequality in (\ref{eq-gauss_approx_high}) captures an alternative viewpoint, which is roughly
the minimum value of $M$ required for H-ARQ to provide a significant advantage.

\begin{figure}
\begin{center}
\subfigure[$\snr=10$ dB]{\label{fig:IRvsR1}
\includegraphics[width = 3.1in]{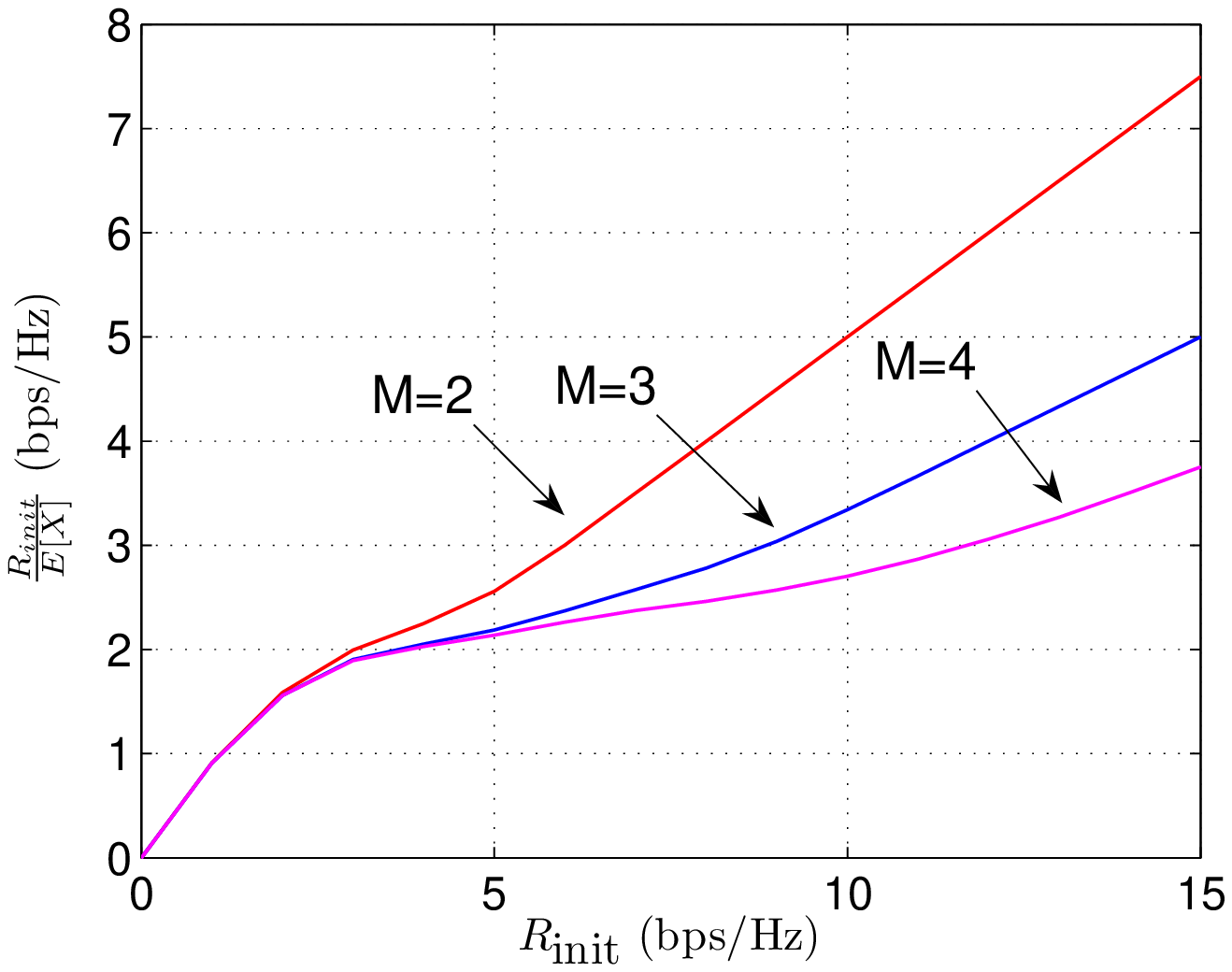}}
\subfigure[$\snr=30$ dB]{\label{fig:IRvsR2}
\includegraphics[width = 3.1in]{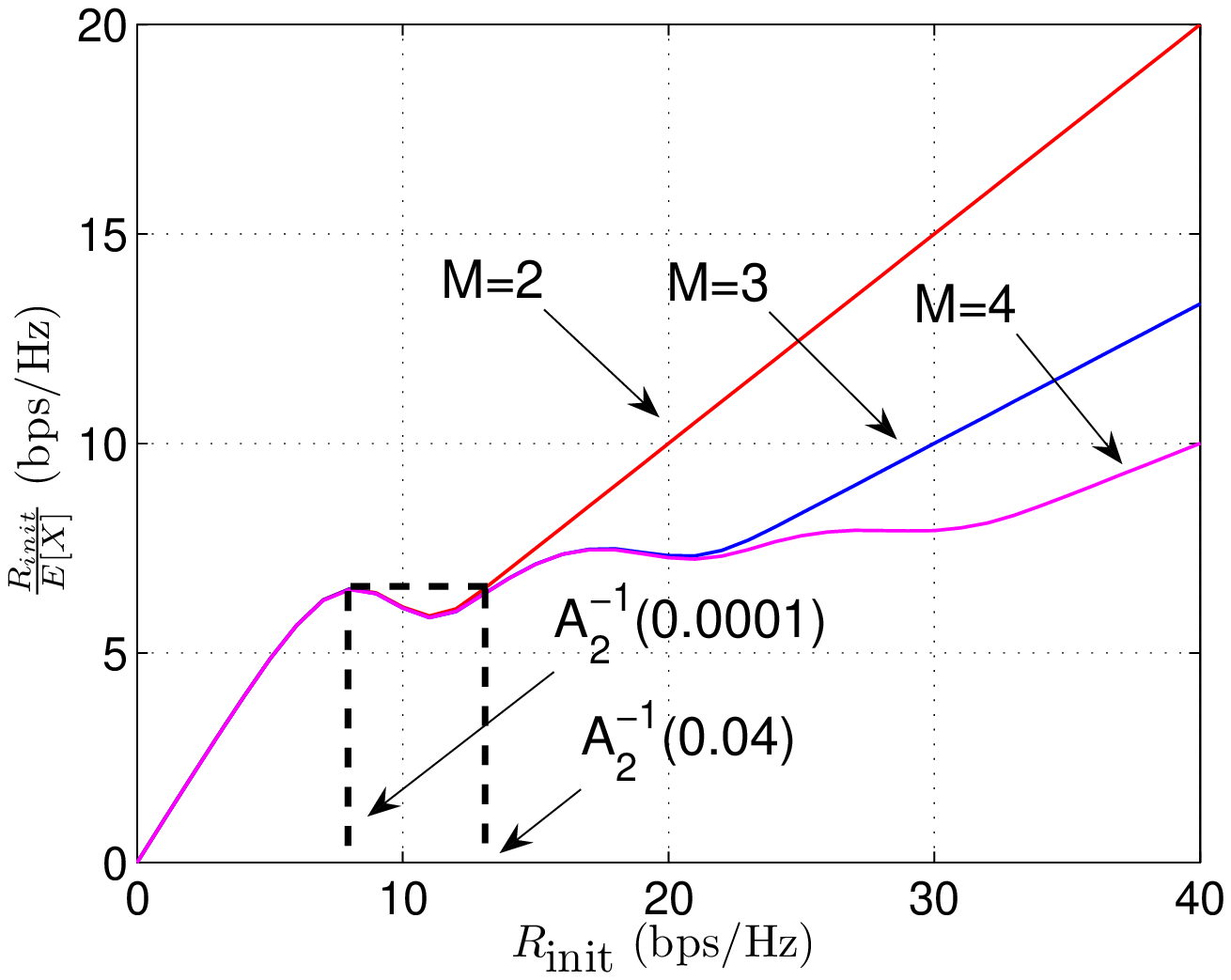}}
\caption{H-ARQ rate $\frac{R_{\textrm{init}}}{\mathbb{E}[X]}$ (bps/Hz) versus initial
rate $R_{\textrm{init}}$ (bps/Hz)} \label{fig:IRvsR}
\end{center}
\end{figure}

Motivated by naive intuition that the H-ARQ rate is monotonically
increasing in the initial rate $R_{\textrm{init}}$, up to this point
we have chosen $R_{\textrm{init}} = M C_{\epsilon}^M =
A_M^{-1}(\epsilon)$ such that outage at H-ARQ termination is exactly
$\epsilon$. However, it turns out that the H-ARQ rate is not always
monotonic in $R_{\textrm{init}}$.
In Fig. \ref{fig:IRvsR}, the H-ARQ rate
$\frac{R_{\textrm{init}}}{{\mathbb E}[X]}$ is plotted versus initial
rate $R_{\textrm{init}}$ for $M=2, ~3,$ and $4$ for $\snr =10$ dB
(left) and  $30$ dB (right).  At $10$ dB,
$\frac{R_{\textrm{init}}}{{\mathbb E}[X]}$ monotonically increases
with $R_{\textrm{init}}$ and thus there is no advantage to
optimizing the initial rate.  At $30$ dB, however,
$\frac{R_{\textrm{init}}}{{\mathbb E}[X]}$ behaves non-monotonically
with $R_{\textrm{init}}$. We therefore define
$\tilde{C}_{\epsilon}^{\textrm{IR},M} (\snr)$ as the
maximized H-ARQ rate, where the maximization is performed over all values
of initial rate $R_{\textrm{init}}$ such that the outage constraint $\epsilon$ is not violated:
\begin{eqnarray}\label{opt_IR}
\tilde{C}_{\epsilon}^{\textrm{IR},M} (\snr) \triangleq
\max_{R_{\textrm{init}} \leq A_M^{-1}(\epsilon)} &&
\frac{R_{\textrm{init}}}{{\mathbb E}[X]}
\end{eqnarray}
The local maxima seen in Fig. \ref{fig:IRvsR} appear to preclude a
closed form solution to this maximization. Although
optimization of the initial rate provides an advantage over a certain SNR range, the
following theorem shows that it does not provide an improvement in the high-SNR offset:
\begin{theorem} \label{thm-high2}
H-ARQ with an optimized initial rate, i.e., $\tilde{C}_{\epsilon}^{\textrm{IR},M}
(\snr)$,  achieves the same high-SNR offset as unoptimized H-ARQ
$C_{\epsilon}^{\textrm{IR},M} (\snr)$
\begin{eqnarray}
\lim_{\snr \rightarrow \infty}
\left[\tilde{C}_{\epsilon}^{\textrm{IR},M}(\snr)-C_{\epsilon}^{\textrm{IR},M}(\snr)\right]=0
\end{eqnarray}
Furthermore, the only initial rate (ignoring $o(1)$ terms) that achieves the correct offset
is the unoptimized value $R_{\textrm{init}} = M C_{\epsilon}^M $.
\end{theorem}
\begin{proof}
See Appendix D.
\end{proof}
In Fig. \ref{fig:IrOptARQ}, rates with and without optimization of the initial rate
are plotted for $\epsilon = 0.01$ and $M=2, 6$.
For $M=2$ optimization begins to make a difference at the point where the unoptimized curve
 abruptly decreases towards $C_{\epsilon}^M$ around $25$ dB, but this advantage vanishes around
 $55$ dB.  For $M=6$ the advantage of initial rate optimization comes about at a much higher SNR,
 consistent with (\ref{eq-gauss_approx_high}).
Convergence of $\tilde{C}_{\epsilon}^{\textrm{IR},6}(\snr)$ to $C_{\epsilon}^{\textrm{IR},6}(\snr)$ does eventually
occur, but is not visible in the figure.

\subsection{Scaling with Outage Constraint $\epsilon$}\label{sec-scalingeps}
\begin{figure}
\begin{center}
\includegraphics[width = 4.5in]{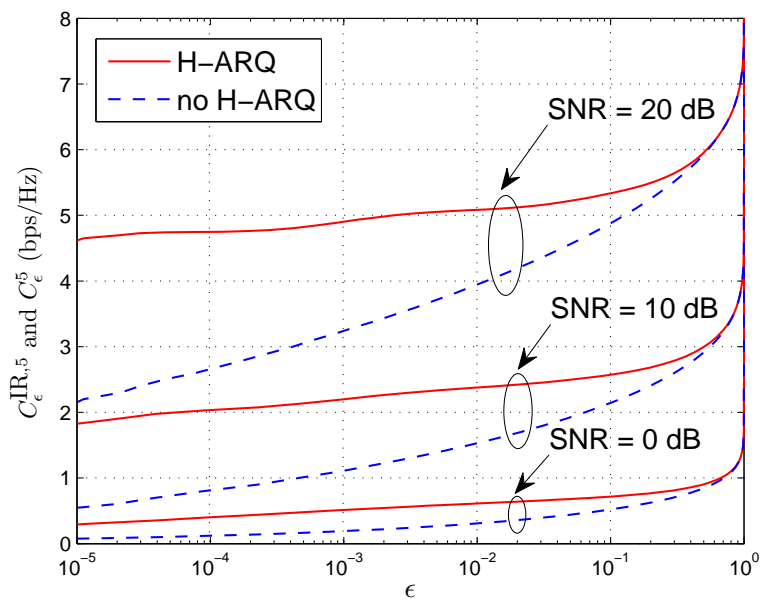}
\caption{H-ARQ rate (bps/Hz) and non-H-ARQ rate (bps/Hz) versus outage probability $\epsilon$ for $M=5$ at $\snr=0$ and $10$ and $20$ dB} \label{fig:R_vs_eps}
\end{center}
\end{figure}

Another advantage of H-ARQ is that the H-ARQ rate is generally less sensitive to the desired outage
probability $\epsilon$ than an equivalent non-H-ARQ system. This advantage is clearly seen in Fig. \ref{fig:R_vs_eps}, where the H-ARQ and non-H-ARQ rates are plotted versus $\epsilon$ for $M=5$ at $\snr=0$, $10$ and $20$ dB.
When $\epsilon$ is large (e.g., roughly around $0.5$) H-ARQ provides almost no advantage: a large
outage corresponds to a large initial rate, which in turn means early termination rarely occurs.
However, for more reasonable values of $\epsilon$, the H-ARQ rate is roughly constant with respect
to $\epsilon$ whereas the non-H-ARQ rate decreases sharply as $\epsilon \rightarrow 0$. The transmitted
rate must be decreased in order to achieve a smaller $\epsilon$ (with or without H-ARQ), but with H-ARQ this
decrease is partially compensated by the accompanying decreasing in the number of rounds $\mathbb{E}[X]$.

\subsection{Chase Combining}\label{sec-chase}
If Chase combining is used, a packet is retransmitted whenever a NACK is received and the receiver performs
maximal-ratio-combining (MRC) on all received packets. As a result,
SNR rather the mutual information is accumulated over H-ARQ rounds and
the outage probability is given by:
\begin{eqnarray}
\Pout^{\textrm{CC},M}(R_{\textrm{init}}) = {\mathbb P}\left[\log_2 \left( 1+\snr\sum_{i=1}^M|h_i|^2 \right)\leq R_{\textrm{init}}
\right].
\end{eqnarray}
For outage $\epsilon$, the initial rate is
$R_{\textrm{init}} = \log_2\left(1 +
F_{\epsilon}^{-1}\left(\sum_{i=1}^M|h_i|^2\right) \snr\right)$.

\begin{figure}
\begin{center}
\includegraphics[width = 4.5in]{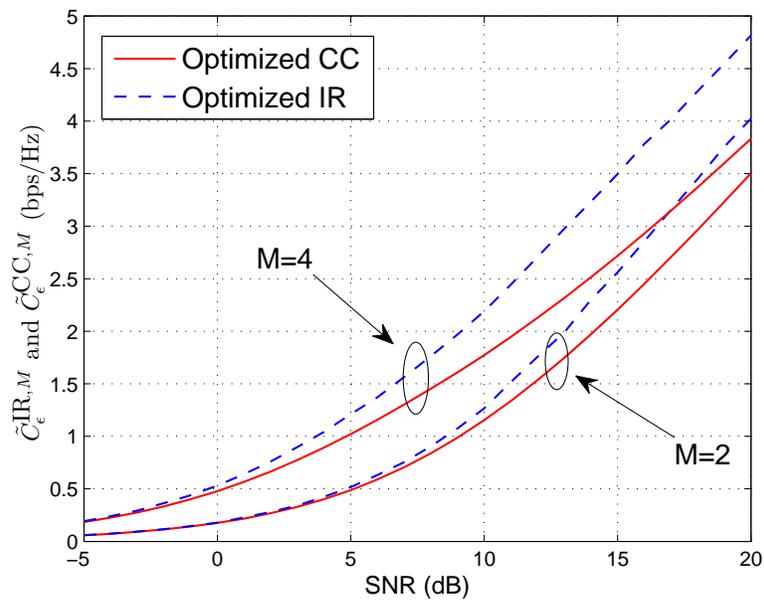}
\caption{Optimized H-ARQ rate $\tilde{C}_{\epsilon}^{\textrm{IR},M}$ (bps/Hz) and $\tilde{C}_{\epsilon}^{\textrm{CC},M}$ (bps/Hz) versus $\snr$ (dB) for $\epsilon = 0.01$}
\label{fig:CCARQnARQ}
\end{center}
\end{figure}

Different from IR, the expected number of H-ARQ
rounds in CC is not dependent on SNR and thus the average rate for outage $\epsilon$ can be written in closed
form:
\begin{equation}\label{CC-outC}
C_{\epsilon}^{\textrm{CC},M}(\snr) = \frac{R_{\textrm{init}}}{{\mathbb E}[X]} = \frac{\log_2\left(1 +
F_{\epsilon}^{-1}\left(\sum_{i=1}^M|h_i|^2\right) \snr\right)}{M -
e^{-F_{\epsilon}^{-1}\left(\sum_{i=1}^M|h_i|^2\right)}
\sum_{k=1}^{M-1}(M-k)\frac{\left(F_{\epsilon}^{-1}\left(\sum_{i=1}^M|h_i|^2\right)\right)^{k-1}}{(k-1)!}},
\end{equation}
where the denominator is ${\mathbb E}[X]$. According to (\ref{CC-outC}), we
can get the high SNR affine approximation as:
\begin{eqnarray}\label{cc-affine}
C_{\epsilon}^{\textrm{CC},M}(\snr) = \frac{1}{{\mathbb E}[X]}\log_2 \snr +
\frac{1}{{\mathbb E}[X]} \log_2
\left(F_{\epsilon}^{-1}\left(\sum_{i=1}^M|h_i|^2\right)\right)+o(1).
\end{eqnarray}
Because ${\mathbb E}[X] > 1$ for any positive outage value, the pre-log factor (i.e., multiplexing gain)
is $\frac{1}{{\mathbb E}[X]}$ and thus is less than one.  This implies that CC performs poorly at high SNR.    This is to be expected because CC is essentially
a repetition code, which is spectrally inefficient at high SNR.  As with IR, the performance
of CC at high SNR can be improved through rate optimization.  At high SNR, the pre-log is critical and thus the initial rate should be selected so
that ${\mathbb E}[X]$ is close to one and thereby avoiding H-ARQ altogether.
Even with optimization, CC is far inferior to IR at moderate and high SNR's.
On the other hand, CC performs reasonably well at low SNR.  This is because $\log(1 + x) \approx x$
for small values of $x$, and thus SNR-accumulation is nearly equivalent to
mutual information-accumulation.
In Fig. \ref{fig:CCARQnARQ} rate-optimized IR and CC are plotted for $M=2, 4$ and $\epsilon=0.01$,
and the results are consistent with the above intuitions.


\section{Conclusion}\label{sec-con}

In this paper we have studied the performance of hybrid-ARQ in the context of an open-loop/fast-fading system
in which the transmission rate is adjusted as a function of the average SNR
such that a target outage probability is not exceeded. The general findings are that H-ARQ provides a significant rate advantage relative to a system not using H-ARQ
at reasonable SNR levels, and that H-ARQ provides a rate quite close to the ergodic capacity even when the channel selectivity
is limited.

There appear to be some potentially interesting extensions of this work.  Contemporary
cellular systems utilize simple ARQ on top of H-ARQ, and
it is not fully understood how to balance these reliability mechanisms; some results in this direction
are presented in \cite{wp09globecom}.  Although we have assumed error-free ACK/NACK feedback, such
errors can be quite important (c.f., \cite{ericsson}) and merit further consideration.  Finally, while we have considered only the mutual information of Gaussian inputs, it is of interest to extend the results to discrete constellations and possibly compare to the performance of actual codes.

%

\section{Acknowledgments}

The authors gratefully acknowledge Prof. Robert Heath for suggesting use of the Gaussian approximation,
and Prof. Gerhard Wunder for suggesting use of Chebyshev's inequality in Section \ref{section-gap}.

\appendices
\section{PROOF OF THEOREM $2$}
Because $C_{\epsilon}^{\textrm{IR},M} = \frac{M}{\mathbb{E}[X]}C_{\epsilon}^M$ and
$\lim_{M\rightarrow\infty}C_{\epsilon}^M = \mu$ (Section \ref{section-gap}), we can prove
$\lim_{M\rightarrow\infty}C_{\epsilon}^{\textrm{IR},M} = \mu$ by showing
$\lim_{M\rightarrow\infty}\frac{M}{\mathbb{E}[X]} = 1$.  Because $\mathbb{E}[X] \leq M$, we can show
this simply by showing $\mathbb{E}[X]$ is of order $M$.
For notational convenience we define $Y_i \triangleq \log_2(1+\snr|h_i|^2)$, and then have:
\begin{eqnarray}
\mathbb{E}[X]&\geq& 1 + \sum_{k=1}^{M-1}\mathbb{P}\left[\sum_{i=1}^k Y_i \leq M\left(\mu - \frac{\sigma}{\sqrt{M\epsilon}}\right)\right] \\
&\geq& 1+ \sum_{k=1}^{\lceil M-M^{\frac{3}{4}}\rceil}\mathbb{P}\left[\sum_{i=1}^k Y_i \leq M \mu - \sigma \sqrt{\frac{M}{\epsilon}}\right]\\
&\geq& 1+ \lceil M-M^{\frac{3}{4}}\rceil \cdot \mathbb{P}\left[\sum_{i=1}^{\lceil M-M^{\frac{3}{4}}\rceil}Y_i \leq M \mu - \sigma \sqrt{\frac{M}{\epsilon}}\right]\label{Ex_low},
\end{eqnarray}
where the first line holds because $\mathbb{E}[X]$ is increasing in $R_{\textrm{init}}$ and
$R_{\text{init}} = M C_{\epsilon}^M \geq M\left(\mu - \frac{\sigma}{\sqrt{M\epsilon}}\right)$ from
(\ref{C_eps_lower}), the second holds because the summands are non-negative, and the last line because the
summands are decreasing in $k$. A direct application of the CLT shows $\mathbb{P}\left[\sum_{i=1}^{\lceil M-M^{\frac{3}{4}}\rceil}Y_i \leq M \mu - \sigma \sqrt{\frac{M}{\epsilon}}\right]\rightarrow 1$ as $M\rightarrow\infty$,
and thus, with some straightforward algebra, we have $\lim_{M\rightarrow\infty}\frac{M}{\mathbb{E}[X]}\rightarrow 1$.

\section{PROOF OF (\ref{app_Ex})}
Firstly, we relax the constraint on $\tilde{X}$ (discreteness and finiteness) to define a new continuous random variable $\hat{X}$, which is distributed along the whole real line. The CDF of $\hat{X}$ (for all real $x$) is
\begin{eqnarray}
\mathbb{P} \left( \hat{X} \leq x \right) = Q \left( \frac{\mu (M - x)-\sqrt{M} \sigma Q^{-1}(\epsilon)}{\sigma \sqrt{x}} \right)\label{xhat}
\end{eqnarray}
Now if we consider the distribution of $\hat{X}-\left(M-\frac{\sigma}{\mu}Q^{-1}(\epsilon)\sqrt{M}\right)$, we have
\begin{eqnarray}
\mathbb{P}\left[\hat{X}-\left(M-\frac{\sigma}{\mu}Q^{-1}(\epsilon)\sqrt{M}\right)\leq x\right]&=&
Q \left( \frac{-\mu x}{\sigma \sqrt{M-\frac{\sigma}{\mu}Q^{-1}(\epsilon)\sqrt{M}+x}} \right)
\end{eqnarray}
where the equality follows from (\ref{xhat}). Notice as $M\rightarrow\infty$,
$\sqrt{M-\frac{\sigma}{\mu}Q^{-1}(\epsilon)\sqrt{M}+x}\rightarrow\sqrt{M}$,
so
\begin{eqnarray}
Q \left(\frac{-\mu x}{\sigma\sqrt{M-\frac{\sigma}{\mu}Q^{-1}(\epsilon)\sqrt{M}+x}}\right)&\rightarrow& Q\left(\frac{-\mu x}{\sigma\sqrt{M}}\right) =  \Phi\left(\frac{x}{\frac{\sigma}{\mu}\sqrt{M}}\right),
\end{eqnarray}
where $\Phi(\cdot)$ is the standard normal CDF with zero mean and
unit variance. So as $M\rightarrow\infty$, the limiting distribution of $\hat{X}$, denoted by $\hat{\Phi}(\cdot)$, goes to $\mathcal{N}\left(M-\frac{\sigma}{\mu}Q^{-1}(\epsilon)\sqrt{M},\left(\frac{\sigma}{\mu}\right)^2
M\right)$, which is
\begin{eqnarray}
\hat{\Phi}(x)=\Phi\left(\frac{x-\left(M-\frac{\sigma}{\mu}Q^{-1}(\epsilon)\sqrt{M}\right)}{\frac{\sigma}{\mu}\sqrt{M}}\right),\quad \quad \quad \quad\text{for $x\in\mathbb{R}$}
\end{eqnarray}
Since ${\mathbb E}[\tilde{X}]$ is an approximation to ${\mathbb E}[X]$,
then we focus on evaluating ${\mathbb E}[\tilde{X}]$ for large $M$:
\begin{eqnarray}
{\mathbb E}[\tilde{X}]& = &\sum_{k=0}^{M-1}\left(1-\mathbb{P}[\tilde{X}\leq k]\right)= M - \sum_{k=1}^{M-1}\tilde{\Phi}(k)\overset{(a)}{=} M - \sum_{k=1}^{M-1}\hat{\Phi}(k)\nonumber\\
& \approx & M- \left(\int_{1}^M\hat{\Phi}(x) dx-\sum_{k=1}^{M-1}\frac{\hat{\Phi}(k+1)-\hat{\Phi}(k)}{2}\right)\nonumber\\
& \overset{(b)}{\approx} &M- \left(\int_1^M\Phi\left(\frac{x-\left(M-\frac{\sigma}{\mu}Q^{-1}(\epsilon)\sqrt{M}\right)}{\frac{\sigma}{\mu}\sqrt{M}}\right) dx-0.5(1-\epsilon)\right)\nonumber\\
&=&M- \int_{M-2\frac{\sigma}{\mu}Q^{-1}(\epsilon)\sqrt{M}}^M\Phi\left(\frac{x-\left(M-\frac{\sigma}{\mu}Q^{-1}(\epsilon)\sqrt{M}\right)}{\frac{\sigma}{\mu}\sqrt{M}}\right)dx-\nonumber\\
&&\quad \
\int_1^{M-2\frac{\sigma}{\mu}Q^{-1}(\epsilon)\sqrt{M}}\Phi\left(\frac{x-\left(M-\frac{\sigma}{\mu}Q^{-1}(\epsilon)\sqrt{M}\right)}{\frac{\sigma}{\mu}\sqrt{M}}\right)dx+0.5(1-\epsilon)\label{dual_int}
\end{eqnarray}
where (a) holds since $\tilde{\Phi}(k)=\hat{\Phi}(k)$ when
$k=1,2,\cdots,M-1$ and (b) follows from
$\hat{\Phi}(M)=1-\epsilon$ and $\hat{\Phi}(1)$ is negligible when
$M$ is large enough. Actually, the first integral in (\ref{dual_int}) can be evaluated as $\frac{\sigma}{\mu}Q^{-1}(\epsilon)\sqrt{M}$
because the expression inside the integral is symmetric with
respect to $M-\frac{\sigma}{\mu}Q^{-1}(\epsilon)\sqrt{M}$ and $\Phi(x)+\Phi(-x)=1$ for any $x\in\mathbb{R}$. For large $M$, the second
integral in (\ref{dual_int}) can be approximated as:
\begin{eqnarray} 
&&\frac{\sigma}{\mu}\sqrt{M}\int_{Q^{-1}(\epsilon)}^{\frac{\sqrt{M}\mu}{\sigma}}Q\left(x\right)dx
\overset{(a)}{\approx}\frac{\sigma}{\mu}\sqrt{M}\int_{Q^{-1}(\epsilon)}^{\infty}Q(x)dx-\frac{\sigma}{2.5\mu}\sqrt{M}\int_{\frac{\sqrt{M}\mu}{\sigma}}^{\infty}\frac{e^{-\frac{x^2}{2}}}{x}dx\nonumber\\
& = &
\frac{\sigma}{\mu}\sqrt{M}\int_{Q^{-1}(\epsilon)}^{\infty}Q(x)dx-\frac{\sigma}{5\mu}\sqrt{M}E_1\left(\frac{M\mu^2}{2\sigma^2}\right)
\approx \frac{\sigma}{\mu}\sqrt{M}\int_{Q^{-1}(\epsilon)}^{\infty}Q(x)dx£¬
\end{eqnarray}
where (a) follows from \cite{Nick}: when
$x$ is positive and large enough, $Q(x)\approx\frac{e^{-\frac{x^2}{2}}}{2.5x}$.  The last line holds because
$\frac{\sigma}{5\mu}\sqrt{M}E_1\left(\frac{M\mu^2}{2\sigma^2}\right)\approx 0$ when M is sufficiently large \cite{Abramowitz}.
This finally yields:
\begin{eqnarray}
{\mathbb E}[X]\approx {\mathbb E}[\tilde{X}] \approx
M-\frac{\sigma}{\mu}Q^{-1}(\epsilon)\sqrt{M}-\frac{\sigma}{\mu}\sqrt{M}\int_{Q^{-1}(\epsilon)}^{\infty}Q(x)dx+0.5(1-\epsilon).
\end{eqnarray}

\section{PROOF OF THEOREM $3$}
In order to prove the theorem, we first establish the following
lemma:
\begin{lemma} \label{lemma-ex}
If the initial rate $R_{\textrm{init}}$ has a pre-log of $r$, i.e., $\lim_{\snr
\rightarrow \infty} \frac{ R_{\textrm{init}} }{\log_2 \snr} = r$, then
\begin{subnumcases}{\lim_{\snr\rightarrow\infty}\mathbb{P}\left[\sum_{i=1}^k\log_2(1+\snr|h_i|^2)\leq R_{\textrm{init}}\right]=}
1, & for $k<r$  and  $k\in\mathbb{Z}^{+}$\\ \label{lemma1_first}
0, & for $k>r$  and  $k\in\mathbb{Z}^{+}$\label{lemma1_second}
\end{subnumcases}
\end{lemma}
\begin{proof}
For notational convenience we use $\gamma_i$ to denote the quantity
$\snr|h_i|^2$.  We prove the first result by using the fact that
$\sum_{i=1}^k \log_2(1+\gamma_i) \leq k\log_2(1+\max_{i=1,\ldots,k}
\gamma_i)$ which yields:
\begin{eqnarray}
\mathbb{P}\left[\sum_{i=1}^k\log_2(1+\gamma_i)\leq R_{\textrm{init}}\right]&\geq&{\mathbb
P} \left[k\log_2 \left(1+ \max_{i=1,\ldots,k} \gamma_i \right) \leq R_{\textrm{init}}
\right] \nonumber\\
&=& \left( 1 - e^{- \frac{2^{\frac{R_{\textrm{init}}}{k}}-1}{\snr}}
\right)^{k}
 =  \left( 1 - e^{- 2^{ \frac{R_{\textrm{init}}}{k} - \log_2( \snr)}+
\frac{1}{\snr}} \right)^{k},\label{lemma_1}
\end{eqnarray}
where the first equality follows because the $\gamma_i$'s are i.i.d.
exponential with mean $\snr$.  The exponent $\frac{R_{\textrm{init}}}{k} - \log_2
(\snr)$ behaves as $\frac{r-k}{k} \log_2(\snr)$.  If $k < r$ this
exponent goes to infinity.  Because the $\frac{1}{\snr}$ term
vanishes, $e$ is raised to a power converging to $- \infty$, and
thus (\ref{lemma_1}) converges to $1$.  This yields the result in
(\ref{lemma1_first}). To prove (\ref{lemma1_second}) we combine the
property $\sum_{i=1}^k \log_2(1+\gamma_i) \geq
k\log_2(1+\min_{i=1,\ldots,k} \gamma_i)$ with the same argument as above:
\begin{eqnarray}
\mathbb{P}\left[\sum_{i=1}^k\log_2(1+\gamma_i)\leq R_{\textrm{init}}\right] &\leq&
{\mathbb P} \left[k\log_2 \left(1+ \min_{i=1,\ldots,k} \gamma_i
\right) \leq R_{\textrm{init}}
\right] = 1-e^{-k\left(2^{\frac{R_{\textrm{init}}}{k}-\log_2(\snr)}
 - \frac{1}{\snr} \right)}\nonumber
\end{eqnarray}
If $k > r$, $e$ is raised to a power that converges to $0$ and thus
we get (\ref{lemma1_second}).
\end{proof}

We now move on to the proof of the theorem.  Using the expression
for ${\mathbb E}[X]$ in (\ref{ex}) we have:
\begin{eqnarray}
\lim_{\snr\rightarrow\infty}{\mathbb E}[X]&=&\lim_{\snr\rightarrow\infty}1 +
{\mathbb P}\left[\log_2(1+\gamma_1)\leq R_{\textrm{init}} \right]+ \ldots +{\mathbb P}
\left[\sum_{i=1}^{L-1}\log_2 (1+\gamma_i)\leq R_{\textrm{init}} \right]. \label{ex_eq}
\end{eqnarray}
Because $R_{\textrm{init}} = M C_{\epsilon}^M$ has a pre-log of $M$, the lemma
implies that each of the terms converge to one and thus
$\lim_{\snr\rightarrow\infty}{\mathbb E}[X]=M$.

In terms of the high-SNR offset we have:
\begin{eqnarray}
\lim_{\snr\rightarrow \infty} [C_{\epsilon}^{\textrm{IR},M}(\snr) -
C_{\epsilon}^{M}(\snr)]
& = & \lim_{\snr\rightarrow \infty} \left[\frac{R_{\textrm{init}}}{C_{\epsilon}^M(\snr)}-{\mathbb E}[X]\right]\frac{C_{\epsilon}^M(\snr)}{{\mathbb E}[X]}\nonumber\\
&\stackrel{(a)}{\leq} &\lim_{\snr\rightarrow \infty} \left[M-{\mathbb E}[X]\right]C_{\epsilon}^M(\snr) \nonumber\\
& = &\lim_{\snr\rightarrow \infty} \left[M-{\mathbb E}[X]\right](\log_2(\snr)+O(1))\nonumber\\
& \stackrel{(b)}{=} & \lim_{\snr\rightarrow \infty}
\left[M-{\mathbb E}[X]\right]\log_2(\snr), \label{eq_gap1}
\end{eqnarray}
where (a) holds because ${\mathbb E}[X] \geq 1$ and $R_{\textrm{init}} = M
C_{\epsilon}^M(\snr)$, and (b) holds because ${\mathbb E}[X] \rightarrow M$ and
therefore the $O(1)$ term does not effect the limit.

Because the additive terms defining ${\mathbb E}[X]$ in (\ref{ex}) are
decreasing, we lower bound ${\mathbb E}[X]$ as
\begin{eqnarray}
{\mathbb E}[X]&\geq&  M
\mathbb{P}\left[\sum_{i=1}^{M-1}\log_2(1+\gamma_i)\leq R_{\textrm{init}}\right] \geq
M\left( 1 - e^{- 2^{ \frac{R_{\textrm{init}}}{M-1} - \log_2( \snr)} + \frac{1}{\snr}
} \right)^{M-1},
\end{eqnarray}
where the last inequality follows from (\ref{lemma_1}). Plugging this
bound into  (\ref{eq_gap1}) yields:
\begin{eqnarray}
\lim_{\snr\rightarrow \infty}
\left[M-{\mathbb E}[X]\right]\log_2(\snr)
&\leq& \lim_{\snr\rightarrow \infty} M
\left(1 - \left( 1 - e^{- 2^{ \frac{R_{\textrm{init}}}{M-1} - \log_2 \snr} +
\frac{1}{\snr} }
\right)^{M-1} \right) \log_2 (\snr)  \nonumber\\
&=& \lim_{\snr\rightarrow \infty}-M\log_2(\snr) \sum_{j=1}^{M-1}{M-1\choose j}\left( - e^{- 2^{ \frac{R_{\textrm{init}}}{M-1} - \log_2\snr } + \frac{1}{\snr}} \right)^j \nonumber
\end{eqnarray}
where the last line follows from the binomial expansion.
Because $R_{\textrm{init}}$ has a pre-log of $M$, each of the terms is of the form
$\alpha\log_2(\snr) e^{-\snr^\beta}$ for some $\beta
> 0$ and some constant $\alpha$,
and thus the RHS of
the last line is zero. Because $C_{\epsilon}^{\textrm{IR},M}(\snr) \geq
C_{\epsilon}^{M}(\snr)$, this shows $\lim_{\snr\rightarrow \infty}
[C_{\epsilon}^{\textrm{IR},M}(\snr) - C_{\epsilon}^{M}(\snr)]=0$.

\section{PROOF OF THEOREM $4$}

In order to prove that rate optimization does not increase the
high-SNR offset, we need to consider all possible choices of the
initial rate $R_{\textrm{init}}$. We begin by considering all choices of $R_{\textrm{init}}$ with a pre-log of $M$, i.e.,
satisfying $\lim_{\snr \rightarrow \infty} \frac{ R_{\textrm{init}} }{\log_2 \snr} =
M$. Because the proof of convergence of ${\mathbb E}[X]$ in the proof of
Theorem $3$ only requires $R_{\textrm{init}}$ to have a pre-log of $M$, we have
${\mathbb E}[X] \rightarrow M$.  To bound the offset, we write the rate
as $R_{\textrm{init}} = M C_{\epsilon}^{M}(\snr) - f(\snr)$ where $f(\snr)$ is strictly
positive and sub-logarithmic (because the pre-log is $M$), and thus the rate offset is:
\begin{eqnarray}
\frac{M C_{\epsilon}^{M}(\snr) - f(\snr)}{{\mathbb E}[X]} - C_{\epsilon}^{M}(\snr)
& = & \left(M - {\mathbb E}[X] \right) \frac{C_{\epsilon}^{M}(\snr)}{{\mathbb E}[X]}  - \frac{f(\snr)}{{\mathbb E}[X]}.
\end{eqnarray}
By the same argument as in Appendix C, the first term is upper bounded by
zero in the limit.  Therefore:
\begin{eqnarray}
\lim_{\snr \rightarrow \infty} \frac{R_{\textrm{init}}}{{\mathbb E}[X]} - C_{\epsilon}^{M}(\snr)
\leq \lim_{\snr \rightarrow \infty}  - \frac{f(\snr)}{{\mathbb E}[X]}.
\end{eqnarray}
Relative to $C_{\epsilon}^{M}(\snr)$, the offset is either strictly negative (if $f(\snr)$ is bounded) or goes to negative
infinity. In either case a strictly worse offset is achieved.

Let us now consider pre-log factors, denoted by $r$, strictly smaller than $M$ (i.e., $r < M$).
We first consider non-integer values of $r$.  By Lemma \ref{lemma-ex} the first $1 + \lfloor r \rfloor$ terms in the
expression for ${\mathbb E}[X]$ converge to one while the other terms go to zero.  Therefore
${\mathbb E}[X] \rightarrow 1 + \lfloor r \rfloor = \lceil r \rceil$.  The long-term transmitted rate, given by
$\frac{R_{\textrm{init}}}{{\mathbb E}[X]}$, therefore has pre-log equal to $\frac{r}{\lceil r \rceil}$.  This quantity is strictly
smaller than one, and therefore a non-integer $r$ yields average rate with a strictly suboptimal
pre-log factor.

We finally consider integer values of $r$ satisfying $r < M$.
In this case we must separately consider rates of the form $R_{\textrm{init}} = r \log \snr \pm O(1)$ versus those of the
form $R_{\textrm{init}} = r \log \snr \pm o(\log \snr)$.  Here we use $o(\log \snr)$ to denote terms that are sub-logarithmic
and that go to positive infinity; note that we also explicitly denote the sign of the $O(1)$ or $o(\log \snr)$ terms.
We first consider $R_{\textrm{init}} = r \log \snr \pm O(1)$.
By Lemma \ref{lemma-ex} the terms corresponding to $k=0, \ldots, r-1$ in the
expression for ${\mathbb E}[X]$ converge to one, while the terms
corresponding to $k=r+1, \ldots, M-1$ go to one. Furthermore, the term corresponding to $k=r$
converges to a strictly positive constant denoted $\delta$:
\begin{eqnarray}
\delta &=& \lim_{\snr\rightarrow\infty}\mathbb{P}\left[\sum_{i=1}^r\log_2(1+ \snr|h_i|^2)\leq r\log_2(\snr)\pm
O(1)\right]\nonumber\\
&=& \lim_{\snr\rightarrow\infty}\mathbb{P}\left[\sum_{i=1}^r\log_2(\snr|h_i|^2)\leq r\log_2(\snr)\pm
O(1)\right]=\mathbb{P}\left[\sum_{i=1}^r\log_2(|h_i|^2)\leq \pm
O(1)\right] \label{eq-r_term}
\end{eqnarray}
where the second equality follows from \cite{prasad2005moc}. $\delta$ is strictly positive because the support of
$\log_2(|h_i|^2)$, and thus of the sum, is the entire real line.  As a result,
${\mathbb E}[X] \rightarrow r + \delta$, which is strictly larger than $r$.  The pre-log of the average
rate is then $\frac{r}{r + \delta} < 1$, and so this choice of initial rate is also sub-optimal.

If $R_{\textrm{init}} = r \log \snr + o(\log \snr)$ the terms in the ${\mathbb E}[X]$ expression behave largely the same as above except that
the $k=r$ term converges to one because the $O(1)$ term in (\ref{eq-r_term}) is replaced with a quantity
tending to positive infinity.  Therefore ${\mathbb E}[X] \rightarrow r + 1$, which also yields a sub-optimal
pre-log of $\frac{r}{r+1} < 1$.

We are thus finally left with the choice $R_{\textrm{init}} = r \log \snr - o(\log \snr)$.  This is the same as the above
case except that the $k=r$ term converges to zero.  Therefore ${\mathbb E}[X] \rightarrow r$, and thus the achieved
pre-log is one.  In this case we must explicitly consider the rate offset, which is written as:
\begin{eqnarray}
\frac{R_{\textrm{init}}}{{\mathbb E}[X]} - C_{\epsilon}^M &=& \frac{ r \log \snr - o(\log \snr) }{{\mathbb E}[X]}- C_{\epsilon}^M= \frac{ r \log \snr}{{\mathbb E}[X]}- C_{\epsilon}^M - \frac{o(\log \snr) }{{\mathbb E}[X]}.
\end{eqnarray}
Using essentially the same proof as for Theorem $3$, the difference between the first two terms
is upper bounded by zero in the limit of $\snr \rightarrow \infty$.  Thus, the rate offset
goes to negative infinity.

Because we have shown each choice of $R_{\textrm{init}}$ (except $R_{\textrm{init}}=MC_{\epsilon}^M$) achieves either a strictly
sub-optimal pre-log or the correct pre-log but a strictly
negative offset, this proves both parts of the theorem.


\newpage

\begin{figure}
\begin{center}
\includegraphics[width = 4.5in]{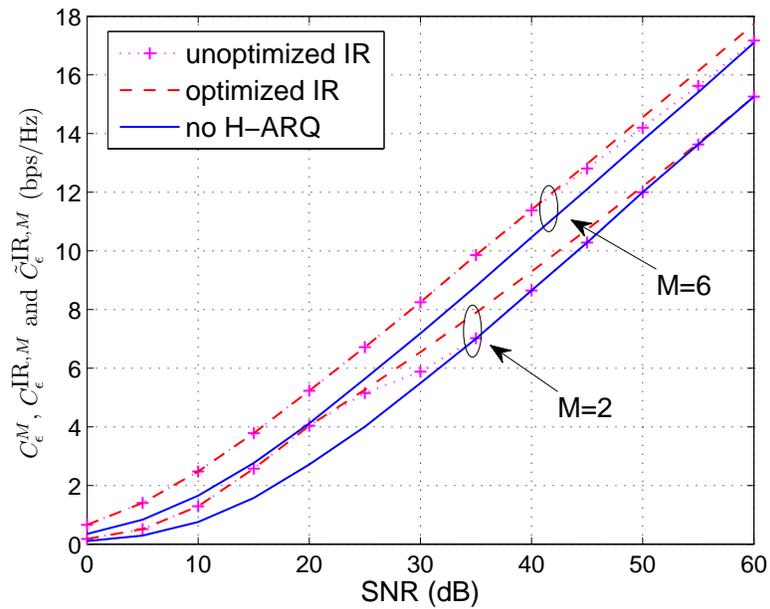}
\caption{H-ARQ rate with/without an optimized initial rate (bps/Hz) and non-H-ARQ rate (bps/Hz) versus $\snr$ (dB) for $\epsilon = 0.01$} \label{fig:IrOptARQ}
\end{center}
\end{figure}

\end{document}